\documentclass[lettersize,journal]{IEEEtran}
\IEEEoverridecommandlockouts
\usepackage{cite}
\usepackage{bm}
\usepackage{amsmath,amssymb,amsfonts}
\usepackage{algorithm}
\usepackage{algpseudocode}
\usepackage{longtable}
\usepackage{graphicx}
\usepackage{amsfonts}
\usepackage{amssymb}
\usepackage{textcomp}
\usepackage[T1]{fontenc}
\usepackage{longtable} 
\usepackage{supertabular} 
\usepackage{xcolor}
\usepackage{subcaption}
\usepackage{tabularx}
\usepackage{comment}
\usepackage{booktabs}
\usepackage{mathrsfs} 
\usepackage{float} 
\usepackage{booktabs} 
\usepackage{mathtools}
\mathtoolsset{centercolon}
\usepackage{multirow} 
\newtheorem{definition}{Definition}

\allowdisplaybreaks
\def\BibTeX{{\rm B\kern-.05em{\sc i\kern-.025em b}\kern-.08em
    T\kern-.1667em\lower.7ex\hbox{E}\kern-.125emX}}

\allowdisplaybreaks
\def\BibTeX{{\rm B\kern-.05em{\sc i\kern-.025em b}\kern-.08em
    T\kern-.1667em\lower.7ex\hbox{E}\kern-.125emX}}
\begin{document}

\title{
Enforcing Cooperative Safety for Reinforcement Learning-based Mixed-Autonomy Platoon Control

\author{Jingyuan Zhou, Longhao Yan, Jinhao Liang, and Kaidi Yang%
\thanks{The authors are with the Department of Civil and Environmental Engineering, National University of Singapore, Singapore 119077. Email:{\{jingyuanzhou, longhao.yan\}@u.nus.edu,
 \{jh.liang, kaidi.yang\}@nus.edu.sg}. (\emph{Corresponding Author: Kaidi Yang})}
 \thanks{This research was supported by the Singapore Ministry of Education (MOE) under its Academic Research Fund Tier 1 (A-8001183-00-00). This article
solely reflects the opinions and conclusions of its authors and not Singapore MOE or any other entity.}
 
 }}

\maketitle

\begin{abstract}
It is recognized that the control of mixed-autonomy platoons comprising connected and automated vehicles (CAVs) and human-driven vehicles (HDVs) can enhance traffic flow. Among existing methods, Multi-Agent Reinforcement Learning (MARL) appears to be a promising control strategy because it can manage complex scenarios in real time. However, current research on MARL-based mixed-autonomy platoon control suffers from several limitations. First, existing MARL approaches address safety by penalizing safety violations in the reward function, thus lacking theoretical safety guarantees due to the black-box nature of RL. Second, few studies have explored the cooperative safety of multi-CAV platoons, where CAVs can be coordinated to further enhance the system-level safety involving the safety of both CAVs and HDVs. Third, existing work tends to make an unrealistic assumption that the behavior of HDVs and CAVs is publicly known and rationale. To bridge the research gaps, we propose a safe MARL framework for mixed-autonomy platoons. Specifically, this framework (i) characterizes cooperative safety by designing a cooperative Control Barrier Function (CBF), enabling CAVs to collaboratively improve the safety of the entire platoon, (ii) provides a safety guarantee to the MARL-based controller by integrating the CBF-based safety constraints into MARL through a differentiable quadratic programming (QP) layer, and (iii) incorporates a conformal prediction module that enables each CAV to estimate the unknown behaviors of the surrounding vehicles with uncertainty qualification. Simulation results show that our proposed control strategy can effectively enhance the system-level safety through CAV cooperation of a mixed-autonomy platoon with a minimal impact on control performance.
\end{abstract}

\section{Introduction}
\IEEEPARstart{C}{onnected} and Automated Vehicles (CAVs) have demonstrated enormous potential to improve traffic flow \cite{deng2023cooperative,liang2022mas,WANG2024104743,tan2024connected}. To realize such potential, significant research attention has been placed on the longitudinal control of CAVs within a platoon of vehicles, primarily aimed at optimizing fuel consumption \cite{zhao2018platoon,yang2022eco}, improving string stability \cite{wang2021leading,shi2021connected,wang2022distributed,shu2023safety,shi2023deep}, addressing privacy issues \cite{zhang2023privacy,ZHOU2024104885}, and enhancing safety \cite{fang2024human,zhou2024enhancing,liang2024enhancing}. 
One widely recognized challenge associated with the longitudinal control of CAVs lies in the prolonged transition period with mixed autonomy, during which CAVs and human-driven vehicles (HDVs) coexist, as technological maturity and social acceptance can only evolve gradually. 
For the control of mixed-autonomy platoons, it is important to coordinate CAVs to improve the performance of the entire platoon, considering the behavior of HDVs.

One promising scheme for mixed-autonomy platoon control is the leading cruise control (LCC) framework \cite{wang2021leading,shang2024decentralized,shi2021connected,li2021reinforcement,shi2023deep,liu2023safety}. Unlike traditional HDVs and CAVs with other longitudinal control schemes (e.g., Adaptive Cruise Control) that only follow their leading vehicles, LCC fundamentally revolutionizes the operation of vehicles by allowing the decision-making process of each CAV to utilize information from both preceding and following vehicles, thereby offering enormous potential to stabilize traffic flow and improve fuel efficiency.
A number of control algorithms have been proposed to exploit such potential, whereby the key is (i) to efficiently coordinate CAVs within a platoon and (ii) to model the unknown and potentially heterogeneous behavior of HDVs. For example, Ref. \cite{wang2022data} proposes a data-driven control scheme for a single CAV to achieve safe and optimal control in mixed traffic using data-enabled predictive control (DeePC), whereby the behavior of HDVs is implicitly captured via a Hankel matrix comprised of historical data~\cite{coulson2019data}. However, DeePC requires solving a quadratic programming problem, which can be computationally expensive if the number of vehicles in the platoon becomes large. 
To reduce computation complexity, Ref. \cite{shang2024decentralized} utilizes decentralized DeePC, whereby each
CAV computes its control input based on locally available data
from its involved subsystem. Nevertheless, the construction of the Hankel matrix still assumes linear system dynamics, which may not be able to represent inherently nonlinear HDV behavior. 

Reinforcement Learning (RL)-based approaches appear to be a promising tool for LCC, which improves computational efficiency and is able to consider nonlinear HDV behavior.
For example, Ref. \cite{shi2021connected} proposes a cooperative RL-based control scheme to enhance string stability and energy-saving performance by coordinating CAVs. 
Ref. \cite{li2021reinforcement} introduces a multi-agent reinforcement learning (MARL) approach, called Communication Proximal Policy Optimization, to enhance driving efficiency in mixed-autonomy platoons while reducing computational complexity. Ref. \cite{shi2023deep} presents a distributed RL-based controller that leverages the aggregated joint behaviors of HDVs to effectively manage the macroscopic features of a mixed-autonomy platoon and improve control efficiency. Compared to model-based controller, such RL-based approaches can more effectively coordinate multiple CAVs in mixed-autonomy platoons by operating in a decentralized manner, where each CAV relies on local observations while being trained in a centralized framework with a shared goal. Additionally, RL-based methods inherently use a data-driven approach to model the unknown behavior of HDVs, making them better suited to adapt to the nonlinear and unknown dynamics of HDVs. 

However, existing MARL-based controllers for mixed-autonomy platoons suffer from three limitations. 
First, existing MARL approaches for mixed-autonomy platoons lack theoretical safety guarantees. The aforementioned works typically consider safety by including a penalty term for safety violations in the reward function for training MARL agents \cite{liu2024reinforcement}, which cannot provide safety guarantees due to the black-box nature of RL. 
Second, few works have exploited the crucial benefits of multi-CAV platoons, i.e., \emph{cooperative safety}, whereby multiple CAVs can be coordinated to enhance \emph{system-level} safety, including their own safety and the safety of other HDVs. 
Third, existing work often assumes that the behavior of HDVs and other CAVs (except for the ego CAV) is explicitly known and rationale \cite{wang2021leading,shi2023deep,zhou2024enhancing}. However, in real traffic scenarios, the behavior of human drivers and other CAVs may be unknown or affected by disturbances due to communication channel interference.
Moreover, as introduced in \cite{zhou2024enhancing}, 
HDVs may exhibit irrational behavior, which may place other vehicles in the platoon in dangerous situations and thereby undermine system-level
safety for both HDVs and CAVs. 

We aim to bridge the aforementioned
research gaps by proposing a safe MARL framework for mixed-autonomy platoons. 
The theoretical foundation of the proposed framework builds on safe RL. 
Despite its rare application in mixed-autonomy platoon control, safe RL has attracted increasing research attention, especially in robotics. 
The main safe RL approaches 
include Constrained Markov Decision Process (CMDP) approaches \cite{chow2018risk}, reachability analysis-based methods \cite{selim2022safe}, and CBF-based approaches \cite{chow2018lyapunov,cheng2019end,emam2021safe,xiao2023barriernet}. In CMDP-based methods, such as \cite{chow2018risk}, primal-dual techniques are used to constrain the MDP. However, CMDP approaches often lack rigorous theoretical guarantees for persistent safety. Reachability analysis-based approaches, e.g.,  \cite{selim2022safe}, combine data-driven reachability analysis with a differentiable collision-checking component to ensure system safety, making them applicable in model-free settings while offering theoretical safety guarantees. Despite these strengths, reachability analysis-based methods face the challenge of the ``curse of dimensionality'', rendering them computationally inefficient, especially in mixed-traffic environments involving multiple CAVs and HDVs. In this paper, we focus on CBF-based approaches thanks to the ease of combining CBF with MARL in the mixed-autonomy platoon control framework. Specifically, CBF can be used to construct a constraint on the control input, which becomes active only when the safety condition is violated, ensuring the algorithm's real-time responsiveness without significantly influencing the effectiveness of the controller. Recently, Ref. \cite{cheng2019end} developed a CBF-quadratic programming (CBF-QP) problem that can compensate for the DRL action such that the compensated action is within a safe set. However, in this work, the CBF-QP is only considered as part of the environment without providing guidance on the training of the DRL model. In contrast, Ref. \cite{xiao2023barriernet} integrated a CBF-QP-based differentiable safety layer into neural network-based controllers using differentiable QP layer~\cite{amos2017optnet}, which can ensure safety while enabling backpropagation of the safety layer and thus facilitating the training of neural networks. 
However, to the best of our knowledge, CBF has rarely been combined with RL in the LCC context. Although our previous work \cite{zhou2024enhancing} extends  \cite{xiao2023barriernet} and provides a safety guarantee by postprocessing RL actions with an optimization model based on CBF, it builds on single-agent RL that only controls a single CAV without considering the coordination of multiple CAVs. 

\emph{Statement of contribution}. 
The contributions of this paper are three-fold. 
First, unlike existing methods that neglect the potential benefits of cooperation in enhancing system-level safety, our approach explicitly considers the notion of \emph{cooperative safety} and characterizes it by a cooperative CBF, enabling CAVs to collaboratively improve the safety of the entire platoon.
Second, we provide a safety guarantee to the MARL-based controller by integrating the CBF-based safety constraints into MARL through a differentiable quadratic programming (QP) layer \cite{amos2017optnet}, which converts the potentially unsafe RL actions to safe actions.
Third, since the cooperation of CAVs involves estimating the decisions of other CAVs and HDVs in the platoon, we devise a conformal prediction module \cite{lindemann2023safe,fontana2023conformal} to explicitly quantify the uncertainty of these estimations and integrate it into the CBF-MARL framework to enhance the robustness against such uncertainties. To the best of our knowledge, the integration between conformal prediction and CBF has been rarely applied to ensure the safety and robustness of mixed-autonomy platoon control.  

The rest of the paper is organized as follows. Section \ref{sec: Preliminaries} introduces the background knowledge about MARL and CBF. Section \ref{sec: Problem Statement} presents the system dynamics and safety concerns of mixed-autonomy traffic. Section \ref{sec: Cooperative Distributed Safe Reinforcement Learning-Based Control Design} provides a detailed description of the proposed cooperative safe MARL-based controller. Section \ref{sec: Simulation Results} performs simulations and analyzes the results. Section \ref{sec: Conclusion} concludes the paper.

\section{Preliminaries}
\label{sec: Preliminaries}
In this section, we introduce Multi-Agent Reinforcement Learning (MARL) and Control Barrier Function (CBF) as theoretical foundations for the proposed approach. 

\subsection{Multi-Agent Reinforcement Learning (MARL)} \label{subsec: MARL}

In this paper, we utilize the decentralized partially observable Markov decision processes (DEC-POMDP) framework, defined by the tuple $\left(\mathcal{X}, \mathcal{U}, \mathcal{O}, \{R^i\}, P, n, \gamma\right)$. Here, $\mathcal{X}$ represents the state space, and $\mathcal{U}$ denotes the shared action space, both of which are the same for each agent~$i$. Agents have partial information about the system states, whereby the local observation made by agent $i$ at global state $\bm{x}^t$ is given by $\bm{o}^t_i = \mathcal{O}(\bm{x}^t; i)$. The transition probability function, $P\left(\bm{x}^{t+1} \mid \bm{x}^t, \bm{u}^t\right)$, describes the probability of the system evolving from state $\bm{x}^t$ to $\bm{x}^{t+1}$ under the joint action $\bm{u}^t = (\bm{u}^t_1, \ldots, \bm{u}^t_n)$ executed by all $n$ agents. The reward function for agent $i$ is represented by $R_i$, and $\gamma \in (0,1]$ is a discount factor representing the weight given to future rewards.

Each agent $i$ employs a policy $\pi_{\bm{\theta}_{\text{RL},i}}(\bm{u}^t_i \mid \bm{o}^t_i)$, parameterized by $\bm{\theta}_{\text{RL},i}$, to determine an action $\bm{u}^t_i$ based on its local observation $\bm{o}^t_i$. This policy is optimized according to the objective: 
$$J\left(\pi_{\bm{\theta}_{\text{RL},i}}\right) = \mathbb{E}_{\bm{o}^t,\bm{u}_i^t\sim \pi_{\bm{\theta}_{\text{RL},i}}}\left[\sum_{t=0}^T\gamma^{t}r(\bm{o}^t,\bm{u}_i^t)\right],$$
which aims to maximize the expected discounted cumulative rewards over the decision horizon.

In this paper, Multi-Agent Proximal Policy Optimization 
(MAPPO) \cite{yu2022surprising} is utilized as the fundamental RL algorithm to train the policies for CAVs. MAPPO has an actor-critic structure, whereby an actor network $\pi_{\bm{\theta}_{\text{RL},i}}$ produces actions based on the observed state $\bm{o}^t_i$, and a shared critic network $V_{\bm{\phi}}(\bm{x}_t)$ estimates the global state-action value of the MDP and generates loss values during agent training. This design enables MAPPO to operate within a Centralized Training and Decentralized Execution (CTDE) framework. In this setup, the actor-critic model learns a centralized critic that processes global information, while simultaneously training a decentralized actor that makes decisions based on local information. This approach effectively balances the benefits of centralized knowledge and decentralized operation for multi-agent systems.

During the training process, the critic network updates its parameters with the aim of minimizing the error between the predicted value function and the actual return:
\begin{equation}
\mathcal{L}(\bm{\phi}_i)=\mathbb{E}_{\bm{x}^t, \bm{u}_i^t \sim \pi_{\bm{\theta}_{\text{RL},i}}}\left[\delta_i^2\right], \quad \delta_i=r_i^t+\gamma V_{\bm{\phi}}\left(\bm{x}^{t+1}\right)-V_{\bm{\phi}}(\bm{x}^t)
\end{equation}
where $\mathcal{L}(\bm{\phi}_i)$ represents the loss function of the critic network, and $\delta_i$ is the temporal-difference error to be approximated. 

The actor network then gets updated following the loss function evaluated by the critic network:
\begin{equation}
\begin{aligned}
&L^{\operatorname{clip}}(\pi_{\bm{\theta}_{i,\text{RL}}})=\\
&\mathbb{E}_{\bm{o}^t_i, \bm{u}_i^t \sim \pi_{i,\bm{\theta}_{\text{RL},\text {old }}}}\left[\operatorname { m i n } \left(\frac{\pi_{\bm{\theta}_{i,\text{RL}}}(\bm{u}^t_i \mid \bm{o}^t_i)}{\pi_{\bm{\theta}_{i,\text{RL},\text {old }}}(\bm{u}_i^t \mid \bm{o}_i^t)} A^{\pi_{\bm{\theta}_{i,\text{RL},\text {old }}}}(
\bm{o}_i^t, \bm{u}_i^t), \right.\right.
\\&\left.\left.\operatorname{clip}\left(\frac{\pi_{\bm{\theta}_{i,\text{RL}}}(\bm{u}_i^t \mid \bm{o}_i^t)}{\pi_{\bm{\theta}_{i,\text{RL},\text {old }}}(\bm{u}_i^t \mid \bm{o}_i^t)}, 1-\epsilon, 1+\epsilon\right) A^{\pi_{\bm{\theta}_{i,\text{RL},\text {old }}}}(\bm{o}_i^t, \bm{u}_i^t)\right)\right]
\end{aligned}
\end{equation}
where the function $\text{clip}(\cdot)$ prevents aggressive updating by utilizing a trust region defined by a hyperparameter $\epsilon$. $ A^{\pi_{i,\bm{\theta}_{\text{RL},\text {old }}}}(\bm{o}_i^t, \bm{u}_i^t)$ is the advantage function that evaluates the benefit of selected actions.

\subsection{Control Barrier Functions}\label{subsec: CBF}
Next, we introduce the concept of control barrier functions (CBFs), which will be later used to characterize the safety conditions of mixed-autonomy platoon control. Let us consider the following control affine system:
\begin{equation}
    \dot{\bm{x}}=f(\bm{x})+g(\bm{x})\bm{u},
    \label{eq: affine control system}
\end{equation}
where $\bm{x} \in \mathcal{D} \subset \mathbb{R}^{n}$ and  $\bm{u} \in \mathcal{U} \subset \mathbb{R}^{m}$. We assume that functions $f: \mathbb{R}^{n} \rightarrow \mathbb{R}^{n}$ and $g: \mathbb{R}^{n} \rightarrow \mathbb{R}^{n\times m}$ are locally Lipschitz continuous. 

Then, CBF is defined in Definition~\ref{def: cbf}. 

\begin{definition}[Control Barrier Function~\cite{ames2014control}]\label{def: cbf}
Let $\mathcal{D} \subset \mathbb{R}^{n}$ be the domain of states. Let the safe set $\mathcal{C}=\left\{\bm{x} \in \mathcal{D} \subset \mathbb{R}^{n}: h(\bm{x}) \geq 0\right\}$ be represented as the superlevel set of a continuously differentiable function $h:\mathcal{D} \rightarrow \mathbb{R}$. Then, $h$ is said to be a control barrier function for system \eqref{eq: affine control system} if there exists an extended class $\mathcal{K}_\infty$  function $\alpha_{\text{CBF}}$ (i.e., a continuous strictly increasing function satisfying $\alpha_{\text{CBF}}(0) = 0$) such that:
\begin{equation}
    \sup _{u \in U}\left[L_{f} h(\bm{x})+L_{g} h(\bm{x}) u\right] \geq-\alpha_{\text{CBF}}(h(\bm{x})), \forall \bm{x}\in \mathcal{D}
    \label{eq: CBF definition}
\end{equation}
where $\displaystyle L_{f}h(\bm{x}):=\Big(\frac{\partial h(\bm{x})}{\partial \bm{x}}\Big)^Tf(\bm{x})$ and $\displaystyle L_{g}h(\bm{x}):=\Big(\frac{\partial h(\bm{x})}{\partial \bm{x}}\Big)^Tg(\bm{x})$ represent the Lie derivatives of CBF candidate $h(\bm{x})$ along system dynamics $f(\bm{x})$ and $g(\bm{x})$. 
\end{definition}

Let us briefly explain why the condition Eq. \eqref{eq: CBF definition} for CBF $h$ can provide a safety guarantee. We can deduce from Eq. \eqref{eq: CBF definition} that there exists control input $u$ such that $\dot{h} = \frac{\partial h(\bm{x})}{\partial \bm{x}}\dot{\bm{x}} = L_{f} h(\bm{x})+L_{g} h(\bm{x}) \bm{u} \geq -\alpha_{\text{CBF}}(h)$. In other words, if the state reaches the boundary of the safe set $\mathcal{C}$ such that $h(\bm{x})=0$, the control input can move the state towards safer states that remain inside the safe set $\mathcal{C}$, i.e., $h(\bm{x})\geq 0$. Therefore, if the system's current state is safe and the control input adheres to the constraints imposed by the CBF, the subsequent states are guaranteed to remain within the safe set~$\mathcal{C}$. Such a property is termed \emph{forward-invariance} \cite{aubin2011viability}. 


\section{Problem Statement}
\label{sec: Problem Statement}
In this section, we present the problem statement, including the modeling of mixed-autonomy platoons in Section \ref{subsec: Mixed Traffic Modelling and Controller Design} and the safety concerns associated with mixed-autonomy platoon control in Section \ref{subsec: safety concerns}.

\subsection{Mixed-Autonomy Platoons Environment Modeling}
\label{subsec: Mixed Traffic Modelling and Controller Design}
\begin{figure*}[h]
    \centering
    \includegraphics[width = 15cm]{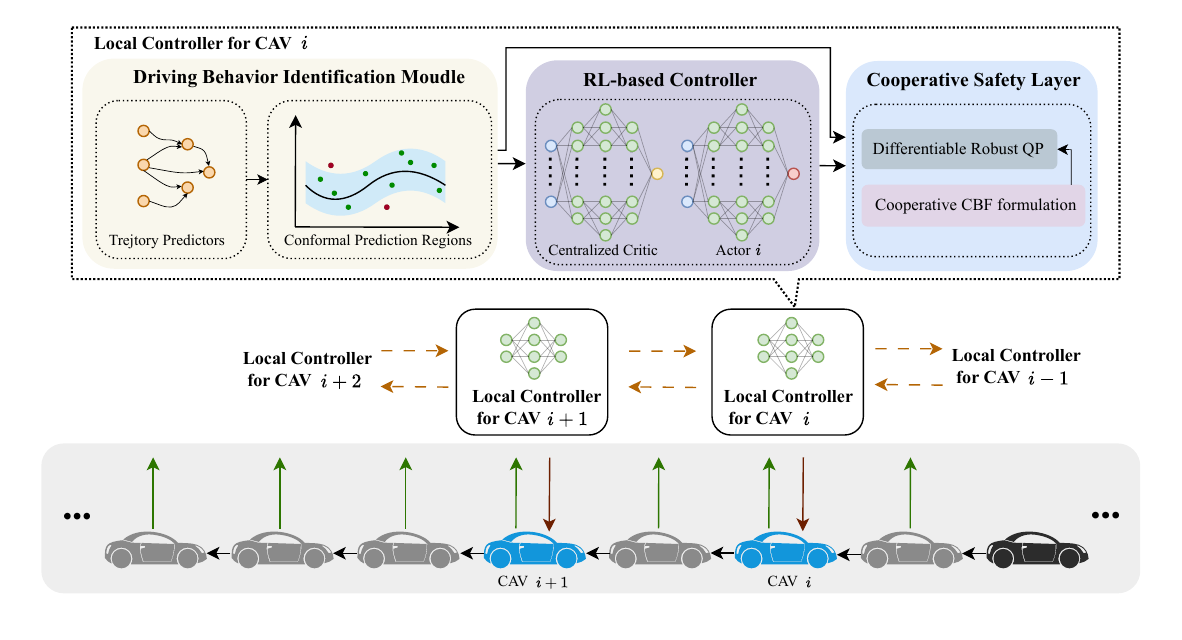}
    \caption{Methodological framework for our proposed controller designed for multiple CAVs in a mixed-autonomy platoon, where blue vehicles represent CAVs, grey vehicles denote HDVs, and the black vehicle indicates the head vehicle.}
    \label{fig:CAV_overview}
\end{figure*}

Let us consider a mixed-autonomy platoon operating in one lane of a freeway segment, which consists of a set of CAVs, denoted by $\Omega_{\mathcal{C}}$, and a set of HDVs, denoted by $\Omega_{\mathcal{H}}$. Let $n=|\Omega_{\mathcal{C}}\cup \Omega_{\mathcal{H}}|$ and $m=|\Omega_{\mathcal{C}}|$ represent the number of vehicles (including CAVs and HDVs) and the number of CAVs, respectively, where $|\cdot|$ denotes the cardinality of a set. Without loss of generality, we consider the scenario with multiple CAVs, i.e., $m\geq 2$. 

The dynamics of these vehicles are described by second-order ordinary differentiable equations as follows: 
\begin{align}
    &\dot{s}_{i}(t) =v_{i-1}(t)-v_{i}(t), \quad i \in \Omega\label{eq:s}\\
    &\dot{v}_{i}(t) =  \left\{\begin{aligned}
        &u_i(t), \quad &i \in \Omega_\mathcal{C} \\
        & \mathbb{F}_i\left(s_i(t), v_i(t), v_{i-1}(t)\right), \quad &i \in \Omega_\mathcal{H}
    \end{aligned}\right. 
     \label{eq:v}
\end{align}
where the states for vehicle $i$ include its speed $v_i$ and the spacing $s_i$ between its preceding vehicle $i-1$. The acceleration rate of HDV $i$ is determined by a car-following model $\mathbb{F}_i(\cdot)$ as a function of the spacing $s_i(t)$, the velocity of the preceding vehicle $v_{i-1}(t)$, and its own velocity $v_i(t)$.  Note that unlike existing works \cite{wang2021leading} that assume the model $\mathbb{F}_i(\cdot)$ to be public knowledge, we make a realistic assumption that both the formulation and parameters of $\mathbb{F}_i(\cdot)$ remain unknown to the CAVs, and each CAV can only learn $\mathbb{F}_i(\cdot)$ from its observations.
The acceleration of a CAV is the control input $u_i(t)$, which will be determined by the MARL controller. Here, following \cite{zhou2024enhancing}, we assume that all vehicles (including HDVs) are connected and can share their state information with other vehicles within the communication range via vehicle-to-vehicle (V2V) communication. Such an assumption will be relaxed in future work. 
Then, $u_i(t)$ is calculated based on the state information of both the preceding and following vehicles of CAV $i$ (denoted by sets $\Omega_{\mathcal{P},i}$  and $\Omega_{\mathcal{F},i}$, respectively) within the communication range. All preceding and following vehicles of CAV $i$ are denoted as $\Omega_{\mathcal{P}_{\text{all}},i}$ and $\Omega_{\mathcal{F}_{\text{all}},i}$, respectively.

We can then rewrite the longitudinal dynamics of the mixed-autonomy platoon \eqref{eq:s}-\eqref{eq:v} as the matrix form: 
\begin{equation}
    \dot{\bm{x}}(t)=f(\bm{x}(t),v_0(t))+B \bm{u}(t),
    \label{eq:continuous system}
\end{equation}
where $\bm{x}(t)=\left[ s_{1}(t), v_{1}(t), \ldots, s_{n}(t), v_{n}(t)\right]^{\top} \in \mathbb{R}^{2n}$ denotes the system states of all vehicles, including CAVs and HDVs. Function $f\left(\cdot\right)$ indicates the vehicle dynamics of both HDVs and CAVs, with $v_0(t)$ indicating the velocity of the head vehicle. The system matrix $B$ for CAVs' control input is summarized as $
    B=\left[\bm{e}_{2 n}^{2 i_1}, \bm{e}_{2 n}^{2 i_2}, \ldots, \bm{e}_{2 n}^{2 i_m}\right] \in \mathbb{R}^{2 n \times m}$, and the vector $\bm{e}^i_{2n}\in\mathbb{R}^{2n}$ associated with CAV $i$ is a vector with the $2i$-th entry being 1 and the others being 0. 

\subsection{Safety Concerns Associated with Mixed-Autonomy Platoon}
\label{subsec: safety concerns}
From the system dynamics described in Section \ref{subsec: Mixed Traffic Modelling and Controller Design}, we can identify two main safety concerns. The first concern is the ego vehicle safety of the CAV, which arises from potential emergency braking by the preceding vehicle, potentially leading to safety-critical events or even a collision between the ego CAV and its preceding vehicle. The second concern is platoon safety, where the following HDVs may behave unpredictably or irrationally, potentially accelerating and causing a collision with their preceding vehicle. These two types of safety concerns are collectively defined as \emph{system-level safety} in Definition \ref{def:system-level safety}.

\begin{definition}[System-Level Safety
]
\label{def:system-level safety}
Given a mixed-autonomy platoon consisting of a set $\Omega_{\mathcal{C}}$ of CAVs and a set $\Omega_{\mathcal{H}}$ of HDVs, system-level safety for this platoon is satisfied if all vehicles after the foremost CAV $i_{\text{first}}$ in the platoon are safe, i.e., the vehicle states satisfy $\bm{x}_i(t)\in\mathcal{C}_i=\left\{\bm{x} \in \mathcal{D} \subset \mathbb{R}^{n}: h_i(\bm{x}) \geq 0\right\}, \forall i\geq i_{\text{first}}, \forall t$, where $\mathcal{D}$ is the domain of states, and $h_i(\cdot)$ represents an operator-defined safety criterion for vehicle $i\in\Omega_{\mathcal{C}}\cup \Omega_{\mathcal{H}}$. 

\end{definition}

We make the following remarks regarding system-level safety for mixed-autonomy platoons introduced in Definition~\ref{def:system-level safety}. First, in contrast to most existing works \cite{shi2021connected,shi2023deep}, that focus only on CAV safety, this notion explicitly considers the safety of HDVs. Second, although our previous work \cite{zhou2024enhancing} considers HDV safety, it overlooks the potential for CAVs to cooperate to enhance platoon safety in scenarios where multiple CAVs are present. This cooperative effort of CAVs to enhance system-level safety is what we refer to as \emph{cooperative safety}, whereby each CAV acts to enhance the safety of itself and vehicles within its communication range while considering the action of other CAVs. 
In addition to mitigating the risks of the actions made by CAVs, such cooperation can further allow CAVs to respond to potential irrational behavior of human drivers, such as sudden acceleration, thereby improving the system-level safety of mixed-autonomy platoons.

Next, we devise a cooperative safe MARL algorithm to address cooperative safety.

\section{Cooperative Safe MARL-Based Control Design}
\label{sec: Cooperative Distributed Safe Reinforcement Learning-Based Control Design}
In this section, we present the cooperative safe MARL approach for mixed-autonomy platoons, as shown in Fig. \ref{fig:CAV_overview}. The proposed approach comprises three main components implemented in each CAV: the MARL-based controller, the cooperative safety layers, and the conformal behavior prediction module.  
The MARL-based controller generates a nominal control input for CAVs, where each CAV is controlled by an identical RL agent to improve the platoon performance. 
The cooperative safety layer converts the nominal control input derived from the MARL-based controller into a safe control action to enhance the safety of the platoon. 
The conformal behavior prediction module enables the robust coordination of CAVs in a distributed manner by allowing each CAV to predict the unknown decisions, i.e., acceleration rates, made by surrounding vehicles to facilitate its decision-making.
Unlike traditional point estimation algorithms that seek to estimate a value, our conformal prediction-based module can provide a confidence region to quantify the uncertainties of the estimation. This approach can be integrated into the MARL module and safety layer to enhance the robustness of the proposed method against such uncertainties.
Next, we present the details of these modules.

\subsection{Multi-Agent DRL-based Controller}
We begin by framing the control of multiple CAVs within a mixed-autonomy platoon as a DEC-POMDP, denoted as $\mathcal{M}=\left(\mathcal{X}, \mathcal{U}, \mathcal{O}, R, P, n, \gamma\right)$. The components of this DEC-POMDP are described as follows. The state space $\mathcal{X}$ includes all possible states of the system, i.e., the spacing between vehicles and their velocities. The action space $\mathcal{U}$ represents the set of joint actions (i.e. the acceleration rates) available to all the CAVs in the platoon. The observation function $\bm{o}^t_i = \mathcal{O}(\bm{x}^t; i)$ denotes the observation for agent $i$ at time $t$ based on the state information and the agent index $i$. These observations of a CAV include the states of other vehicles within its communication range (i.e., vehicles in set $\Omega_{\mathcal{P},i}\cup \Omega_{\mathcal{F},i}$). $P$ is the transition probability, which is defined by the system dynamics in Eq. \eqref{eq:continuous system}. The reward function is represented by $R$, with the detailed design explained as follows.

Our reward formulation includes both global and local rewards, represented as follows: 
\begin{equation}
    R = w_{\text{global}}R_{\text{global}} + w_{\text{local}}\sum_{i\in \Omega_{\mathcal{C}}}R_{i,\text{local}},
\label{eq: global reward}
\end{equation}
where $R_{\text{global}}$ represents the global reward, and $R_{i,\text{local}}$ denotes the local reward for CAV $i\in\Omega_{\mathcal{C}}$. Parameters $w_{\text{global}}\geq 0$ and $w_{\text{local}}\geq 0$ are the weighting coefficients satisfying $w_{\text{global}} + w_{\text{local}} = 1$, which are empirically chosen 
to ensure a balanced optimization between system-level and agent-level goals.

The global reward is designed to reflect the string stability of the platoon~\cite{feng2019string,liu2022structural}. 
String stability ensures that disturbances from the head vehicle do not get amplified through the platoon, which is essential for improving the traffic flow.
Mathematically, let $i_{\text{first}}$ be the foremost CAV in the platoon. Since the coordination of CAVs can only impact the 
propagation of the disturbances on vehicle $i_{\text{first}}-1$ (i.e., the immediate preceding vehicle of CAV $i_{\text{first}}$), the global reward seeks to minimize the velocity oscillations caused by such disturbances on the entire platoon: 
\begin{equation}
    R_{\text{global}} = -(v_{i_{\text{first}}}-v_{i_{\text{first}}-1})^2 - \sum_{j \in \Omega_
\mathcal{H}\setminus\Omega_{\mathcal{P},i_{\text{first}}}} k_{j,\text{stability}}(v_j - v_{i_{\text{first}}-1})^2
\end{equation}
where $k_{j,\text{stability}} \leq 1$ represents weighting coefficient for the velocity oscillation of vehicle $j$. In the current setup, all weighting coefficients are assigned a value of $1$, in line with \cite{wang2023deep,zhou2024enhancing}. Future work may focus on tuning the parameters of the reward function, e.g., using inverse RL \cite{wen2023modeling}.

The local reward  $R_{i,\text{local}}$ for each CAV $i\in\Omega_{\mathcal{C}}$ aims to enhance traffic efficiency and safety, characterized by an efficiency reward $r_{\text{efficiency},i}$ and a safety reward $r_{\text{safety},i}$: 
\begin{equation}
    R_{i,\text{local}} = w_{\text{efficiency}}r_{\text{efficiency},i} + w_{\text{safety}}r_{\text{safety},i}
\label{eq: local reward}
\end{equation}
where the weighting coefficients $w_{\text{efficiency}}$ and $w_{\text{safety}}$ are determined by CAV manufacturers/operators to reflect their preferences regarding these control objectives. 

The traffic efficiency $r_{\text{efficiency},i}$ is described as follows~\cite{vogel2003comparison}:
\begin{align}
    r_{\text{efficiency},i}=\left\{\begin{array}{lc}
                                -1, & \mathrm{TH}_i \geq 2.5 \\
                                0, & \text { otherwise }
                                \end{array}\right.
\end{align}
where $\mathrm{TH}_i(t) = \frac{s_i(t)}{v_i(t)}$ serves as an approximation of the headway, which is represented as the reciprocal of the traffic flow. The headway threshold of $2.5$ is established to maintain a high flow on the road section.

The safety reward $r_{\text{safety},i}$ follows the common practice from previous works (e.g., \cite{zhou2024enhancing}) that leverage the time-to-collision (TTC) metric.
\begin{align}
&r_{\text{safety},i}= \begin{cases}\log \left(\frac{\mathrm{TTC}_i}{4}\right), & 0 \leq \mathrm{TTC} \leq 4 \\ 0, & \text { otherwise }\end{cases}
\label{eq:reward_safety}
\end{align}
where TTC is calculated as $\mathrm{TTC}_i(t)=-\frac{s_i(t)}{v_{i-1}(t)-v_i(t)}$. A negative reward is incurred for unsafe driving conditions with TTC below a threshold of $4$ seconds as in \cite{das2019defining}. 

The proposed MARL-based controllers are trained in a decentralized actor-critic manner using MAPPO (as described in Section \ref{subsec: MARL}). Specifically, we use decentralized actors and a centralized critic to enable centralized training and decentralized execution. The centralized critic enhances collaboration by providing agents with a shared evaluation based on global information during training, promoting coordinated behavior. Decentralized actors reduce computational complexity by making independent decisions during execution, eliminating the need for real-time coordination.

\subsection{Cooperative Control Barrier Function}
\label{subsec: Cooperative Control Barrier Function Design}
Although the MARL agent considers safety by penalizing safety violations in the reward function (see Eq. \eqref{eq:reward_safety}), safety cannot be guaranteed due to the black box nature of MARL. 
To provide a safety guarantee, we aim to design a cooperative safety layer to convert the nominal control input provided by MARL into a safe action. 
To this end, we first establish safety constraints to enforce cooperative safety for mixed-autonomy platoons in this section, and Section III-C then formulates a quadratic programming layer to implement such conversion.

We define the safety requirement of vehicle $i\in \Omega_{\mathcal{C}} \cup \Omega_{\mathcal{H}}$ as the need to maintain sufficient headway from the preceding vehicle, i.e., $    s_i-\tau v_i\geq 0$, where $\tau$ is the desired time headway. Hence, we define the CBF candidate for vehicle $i$ as 
\begin{align}
    h_i = s_i-\tau v_i,~i\in \Omega_{\mathcal{C}} \cup \Omega_{\mathcal{H}}.
\end{align}
When $h_i\geq 0$, we can say that vehicle $i$ is considered safe with respect to its preceding vehicle. 

Based on the CBF candidates, we can represent system-level safety described in Definition~\ref{def:system-level safety} as $h_i \geq 0, \forall i \in \Omega_{\mathcal{C}} \cup \Omega_{\mathcal{H}}$, which involves the safety requirement $h_i\geq 0$ of both CAVs and HDVs. Notice that these safety requirements do not explicitly involve CAV decisions and hence may not directly provide constraints on these decisions. To address this issue, we next convert these safety requirements into safety constraints that explicitly include CAV decisions. 

\vspace{0.2em}\noindent\emph{CAV safety constraints:} 
 By Definition~\ref{def: cbf}, given CBF candidate $h_j$, we can construct the safety constraint for each CAV $j\in\Omega_\mathcal{C}$ as follows:
\begin{align}
    L_fh_j(\bm{x}) + L_gh_j(\bm{x})\bm{u}+\alpha_{j,\text{CBF}}(h_j(\bm{x}))\geq 0,
    \label{eq:cbf_CAV}
\end{align}
where $\alpha_{\text{CBF},j}$ is chosen as a linear function with a positive constant $\gamma_{j,\text{CBF}}$. We can further calculate $L_fh_j(\bm{x}) = v_{j-1} - v_j$, and $L_gh_j(\bm{x})\bm{u} = -\tau u_j$. To better tailor the derived CBF constraint to the DRL settings, we equivalently rewrite Eq.~\eqref{eq:cbf_CAV} as follows:
\begin{equation}
    v_{j-1} - v_j -\tau  (u_{\text{safe},j} + u_{\text{RL},j}) + \gamma_{j,\text{CAV}}(s_j-\tau u_j) \geq 0, j\in\Omega_{\mathcal{C}},
    \label{eq:cbf_CAV_expand}
\end{equation}
where $u_j = u_{\text{safe},j} + u_{\text{RL},j}$ with $u_{\text{RL},j}$ indicating the nominal action from the MARL-based controllers, and $u_{\text{safe},j}$ the compensation calculated from the cooperative safety layer.

\vspace{0.2em}\noindent\emph{HDV safety constraints:} We now establish the safety requirement for the HDV $i\in\Omega_
\mathcal{H}\setminus\Omega_{\mathcal{P},i_{\text{first}}}$. Here, we exclude the HDVs ahead of the first CAV $i_{\text{first}}$ in the platoon, as these HDVs cannot be influenced by the control decisions of CAVs (i.e., their safety cannot be enhanced by controlling CAVs). 

For each HDV $i\in\Omega_{\mathcal{H}}$, one traditional way to construct the safety constraints is to directly apply the conditions in Definition~\ref{def: cbf} and yield: 
\begin{align}
    L_fh_i(\bm{x})  + \gamma_{i,\text{HDV}}h_i(\bm{x})\geq 0, 
    \label{eq:cbf_HDV1}
\end{align}
where the term $L_{g}h_i(\bm{x})\bm{u} = 0$, as the system dynamics of HDVs do not explicitly contain the control decisions of CAVs. Hence, the condition Eq. \eqref{eq:cbf_HDV1} is independent of the control decisions and cannot be used as a constraint to convert the nominal control input. Admittedly, this issue can be addressed by applying high-order CBFs \cite{xiao2021high} that treat \eqref{eq:cbf_HDV1} as another CBF candidate and apply Definition~\ref{def: cbf} until the established conditions become dependent on the control actions. However, the determination of high-order CBFs can be computationally inefficient or even infeasible in practice when applied to mixed-autonomy platoons \cite{zhou2022safety}. 

To address this issue, we propose an alternative approach and define reduced-order CBF candidates by extending our previous work \cite{zhou2024enhancing} to cooperative scenarios. The reduced-order CBF candidates read as follows:
\begin{equation}
    h_{i,\text{suf}} = h_{i} - \sum_{j\in\Omega_{\mathcal{S},i}} k_{i,j,\text{CBF}}h_{j}.   \label{eq:cbf_reduced}
\end{equation}
where the set of preceding CAVs for HDV $i$ within the communication range is given by $\Omega_{\mathcal{S},i} = \Omega_{\mathcal{P},i} \cap \Omega_{\mathcal{C}}$.
Notice that $h_{i,\text{suf}} \geq 0$ is a sufficient condition for $h_i \geq 0$, as we have $h_j \geq 0,~j \in \Omega_{\mathcal{S},i}$ ensured by Eq. \eqref{eq:cbf_CAV_expand} for all CAVs. The key insight of the cooperative CBF design is that in an interconnected system, CAVs positioned ahead of the HDV can cooperatively increase the spacing for the HDV, thereby making its safety constraints more feasible.

The reduced-order CBF candidate defined in Eq. \eqref{eq:cbf_reduced} has a relative degree of 1, because according to Definition~\ref{def: cbf}, the safety constraints associated with $h_{i,\text{suf}}$ depends on the control actions of CAVs in set $\Omega_{\mathcal{S},i}$, represented as follows: 
\begin{align}
    L_{f} h_{i,\text{suf}}(\bm{x})+L_{g} h_{i,\text{suf}}(\bm{x}) \bm{u} + \gamma_{i,\text{HDV}}h_{i,\text{suf}} \geq 0 \label{eq:cbf_HDV_nominal}
\end{align}
where the first two terms can be represented as 
\begin{align}
    &L_{f}h_{i,\text{suf}}(\bm{x}) = L_{f}h_i(\bm{x}) - \sum_{j\in\Omega_{\mathcal{S},i}}k_{i,j,\text{CBF}}L_{f}h_j(\bm{x}) \notag \\
    &= v_{i-1} - v_i - \tau\hat{\mathbb{F}}_i(s_i,v_i,v_{i-1}) -  \sum_{j\in\Omega_{\mathcal{S},i}}k_{i,j,\text{CBF}}\Big(v_{j-1} - v_j\Big),\\
   & L_{g} h_{i,\text{suf}}(\bm{x}) \bm{u}  = L_{g}h_i(\bm{x}) \bm{u} - \sum_{j\in\Omega_{\mathcal{S},i}}k_{i,j,\text{CBF}}L_{g}h_j(\bm{x}) \bm{u} \notag \\ 
    & = -\sum_{j\in\Omega_{\mathcal{S},i}}\tau k_{i,j,\text{CBF}}u_j, 
\end{align} 
where we can calculate $L_fh_i(\bm{x}) = v_{i-1} - v_i -\tau\hat{\mathbb{F}}_i(s_i,v_i,v_{i-1}),~L_gh_i(\bm{x})\bm{u} = 0$ with $\hat{\mathbb{F}}_i$ indicating the estimated car-following model output for HDV $i$. We can also obtain $L_fh_j(\bm{x}) = v_{j-1} - v_j,~L_gh_j(\bm{x})\bm{u} = -\tau u_j,~j\in\Omega_{\mathcal{S},i}$.  


\subsection{Conformal Behavior Prediction Module}
Note that the CBF constraints in Eq. \eqref{eq:cbf_HDV_nominal} require explicit values for the acceleration of surrounding vehicles, which can be estimated using deep learning approaches. 
However, due to the complex and stochastic behaviors of other vehicles in real traffic scenarios, estimation errors are unavoidable and can compromise the safety guarantees provided by CBFs. To address this
 issue, we incorporate a conformal prediction module to enhance the robustness of the proposed method.

The conformal behavior prediction module allows each CAV $j\in\Omega_{C}$ to predict the vehicle acceleration $\hat{a}_i^{t|t-1}$ with probabilistic error bounds of every other vehicle $i \in \Omega_{\mathcal{P},j}\cup \Omega_{\mathcal{F},j}$ within the communication range in the platoon, using the observations $y_i^{t-1}$ about vehicle $i$ at the previous time step $t-1$.  
The module includes two steps. First, we develop a deterministic neural network-based predictor to predict the vehicle acceleration rate of each vehicle. Second, we leverage conformal prediction to quantify the uncertainty of the acceleration predictions in real time. Specifically, we calculate a confidence region corresponding to a predefined failure probability $\epsilon_{\text{con}}$, such that the true acceleration rates fall within this confidence region with a probability of $1-\epsilon_{\text{con}}$. Further details on the module are provided next.

\subsubsection{Acceleration Rate Predictors}
\label{subsubsec:Trajectory Predictors}
In the first step, to map the observations $y_i^{t-1}$ to the acceleration at the current time step $t$, $\hat{a}_i^{t\mid t-1}$, we first collect a dataset $\mathcal{D}_i$ of comprising historical observations and ground-truth acceleration and split it into training dataset $\mathcal{D}_{i,\text{train}}$ and calibration dataset $\mathcal{D}_{i,\text{cal}}$. 
The observations about a CAV include its spacing, velocity, and acceleration, along with the velocity of its preceding vehicle. In contrast, the observations about an HDV are limited to the HDV's spacing and velocity, as well as the velocity of its preceding vehicle, but not the HDV's acceleration. This is because the car-following behavior for HDVs does not significantly rely on acceleration data. Subsequently, we design a fully connected neural network-based predictor $\hat{a}_i^{t\mid t-1}=\xi_i(y_i^{t-1})$ for each vehicle, which is trained using data from $\mathcal{D}_{i,\text{train}}$. 

\subsubsection{Conformal Prediction Regions}
\label{subsubsec:Conformal Prediction Regions}
To quantify the inherent uncertainty in predictions, we define a confidence region corresponding to a given failure probability $\epsilon_{\text{con}} \in\left(0,1\right)$. This confidence region is characterized by a threshold value $C^{t\mid t-1}$ such that
\begin{equation}
    P\left(R^{t\mid t-1}\leq C^{t\mid t-1}\right)\geq 1-\epsilon_{\text{con}},
    \label{eq: probability bound}
\end{equation}
where $R^{t\mid t-1}_i\coloneqq \max_{j\in \Omega_{\mathcal{P},i}\cup \Omega_{\mathcal{F},i}}\|a_j^{t} - \hat{a}_j^{t\mid t-1}\|$ represents the nonconformity score. This score quantifies the discrepancy between the true accelerations sequence of all vehicles, $\{a_j^{t}\}_{j \in \Omega_{\mathcal{P},i}\cup \Omega_{\mathcal{F},i}}$, and the predicted acceleration sequence $\{\hat{a}_j^{t\mid t-1}\}_{j \in \Omega_{\mathcal{P},i}\cup \Omega_{\mathcal{F},i}}$. 

The confidence region, i.e., the specific threshold value $C^{t\mid t-1}$, is determined using the calibration dataset $\mathcal{D}_{\text{cal}} = \cup_{i\in \Omega_{\mathcal{C}}} \mathcal{D}_{i,\text{cal}}$. For each sample $l \in \mathcal{D}_{\text{cal}}$, we compute $R^{t\mid t-1}_{(l)}$. We then sort these nonconformity scores in non-decreasing order. To facilitate the next step in our methodology, we append an additional term $R_{\left(|D_{\text{cal}}|+1\right)}^{t\mid t-1}\coloneqq \infty$ to our sorted list of scores, representing the value indexed in $\left(|D_{\text{cal}}|+1\right)$.

Then, the threshold value $C^{t\mid t-1}$ is determined as the $(1-\epsilon_{\text{con}})$-quantile of the nonconformity scores:
\begin{equation}
    C^{t\mid t-1} \coloneqq R_{(p)}^{t\mid t-1}
    \label{eq: prediction error regions}
\end{equation}
where the index is set as $p = \lceil \left(|D_{\text{cal}}|+1\right)\left(1-\epsilon_{\text{con}}\right) \rceil$, ensuring that the confidence region constructed by the selected threshold $C^{t\mid t-1}$ satisfies the inequality specified in Equation \eqref{eq: probability bound}, thereby delivering a statistically valid measure of uncertainty. 

\subsection{Robust Differentiable Quadratic Programming Layer}

Based on the cooperative CBF constraints and conformal prediction, we formulate a quadratic programming (QP) problem for each CAV to convert the nominal control input from MARL into safe actions. The input to the QP problem includes the nominal control input from MARL and the confidence region as the output of the conformal prediction module. Then, we formulate the QP problem based on the safety constraints. Finally, we integrate the QP problem into the MARL framework as a differentiable QP layer, which allows the joint training of the MARL and QP-based safety layer.

\subsubsection{Combining of Conformal Prediction and Cooperative CBF}

For presentation simplicity, let us represent the proposed CBFs in Section \ref{subsec: CBF} in the general linear form with constant bias with parameters $\boldsymbol{c}_1\in\mathbb{R}^{2n}, \boldsymbol{c}_2\in\mathbb{R}$:  
\begin{equation}
    h = \boldsymbol{c}_1^\top \boldsymbol{x} + \boldsymbol{c}_2. 
\end{equation}

Assuming that the system dynamics and all the CAVs' control inputs are explicitly known, the CBF constraints can be directly specified using Definition~\ref{def: cbf} and Eq. \eqref{eq: CBF definition} as follows:
\begin{equation}
\begin{aligned}
&\boldsymbol{c}_1^\top f(\bm{x}) + \boldsymbol{c}_1^\top g(\bm{x})\boldsymbol{u} + \alpha_{\text{CBF}}(h(\bm{x}))\geq 0.
\end{aligned}
\label{eq: original CBF}
\end{equation}
where $f: \mathbb{R}^{2n} \rightarrow \mathbb{R}^{2n}$, $g: \mathbb{R}^{2n} \rightarrow \mathbb{R}^{2n\times m}$, and $u \in \mathbb{R}^{m}$.

However, the aforementioned CBF does not account for estimation errors in the states of surrounding vehicles. To address this, we incorporate conformal prediction error bounds into the CBF constraints to compute a sufficient condition that ensures the robustness of the safety constraints, given by:
\begin{equation}
\begin{aligned}
&\boldsymbol{c}_1^\top f(\bm{x}) + \boldsymbol{c}_1^\top g(\bm{x})\boldsymbol{u} + \alpha_{\text{CBF}}(h(\bm{x}))\geq E_{\text{con}}.
\end{aligned}
\label{eq: CBF with conformal}
\end{equation}
where $E_{\text{con}}$ defines a safety margin by providing a sufficient condition for the CBF to ensure its robustness in response to these disturbances. The details of $E_{\text{con}}$ are given as follows.

In our setting, we use the predicted information $\hat{f}(\bm{x})=f(\bm{x})+\boldsymbol{e}_{f}$ and $\hat{\boldsymbol{u}}=\boldsymbol{u}+\boldsymbol{e}_{u}$ in our CBF formulation, where $\hat{f}(\bm{x})\in\mathbb{R}^{2n}$ represents the system dynamics with estimated human driver behavior, $\hat{\boldsymbol{u}}\in\mathbb{R}^{m}$ is the vector contains real-time nominal control input for the ego vehicle and the estimated control input for other CAVs, $\boldsymbol{e}_{f}\in\mathbb{R}^{2n}$ and $\boldsymbol{e}_{u}\in\mathbb{R}^{2n}$ are the error between the true values and the predicted values. Then, we can rewrite Eq. \eqref{eq: original CBF} as:
\begin{align}
&\boldsymbol{c}_1^\top (\hat{f}(\bm{x})-\boldsymbol{e}_{f}) + \boldsymbol{c}_1^\top g(\bm{x})(\hat{\boldsymbol{u}}-\boldsymbol{e}_{u}) + \alpha_{\text{CBF}}(h(\bm{x}))\geq 0,
\end{align}
which can be organized into 
\begin{align}
&\boldsymbol{c}_1^\top \hat{f}(\bm{x})+ \boldsymbol{c}_1^\top g(\bm{x})\hat{\boldsymbol{u}} + \alpha_{\text{CBF}}(h(\bm{x}))\geq \boldsymbol{c}_1^\top \boldsymbol{e}_{f} + \boldsymbol{c}_1^\top g(\bm{x})\boldsymbol{e}_{u}. \label{eq: revised CBF}
\end{align}

Although the prediction errors $\boldsymbol{e}_{f}$ and $\boldsymbol{e}_{u}$ are unknown and time-varying, they can be bounded by the conformal prediction bound $C^{t\mid t-1}$ with a probability $(1-\epsilon_{\text{con}})\times 100$\%, resulting in $|\boldsymbol{e}_{f}|\leq C^{t\mid t-1}\bm{1}_{2n},~|\boldsymbol{e}_{u}|\leq C^{t\mid t-1}\bm{1}_{2n}$, where $|\cdot|$ represents the element-wise absolute function of a vector, and $\bm{1}_{2n}\in\mathbb{R}^{2n}$ represents a vector of all 1s. Using these conformal prediction bounds and the Cauchy-Schwarz Inequality, we obtain $\boldsymbol{c}_1^\top \boldsymbol{e}_{f}\leq|\boldsymbol{c}_1^\top||\boldsymbol{e}_{f}|\leq C^{t\mid t-1}|\boldsymbol{c}_1^\top |\bm{1}_{2n}$, $\boldsymbol{c}_1^\top g(\bm{x})\boldsymbol{e}_{u}\leq|\boldsymbol{c}_1^\top||g(\bm{x})||\boldsymbol{e}_{u}|\leq C^{t\mid t-1}|\boldsymbol{c}_1^\top||g(\bm{x})|\bm{1}_{2n}$. Consequently, the CBF constraint in Eq. \eqref{eq: revised CBF} can be sufficiently represented by:
\begin{equation}
\begin{aligned}
&\boldsymbol{c}_1^\top \hat{f}(x) + \boldsymbol{c}_1^\top g(\bm{x})\hat{\boldsymbol{u}} + \alpha_{\text{CBF}}(h(\bm{x}))\geq\\
&C^{t\mid t-1}|\boldsymbol{c}_1^\top|\bm{1}_{2n}  + C^{t\mid t-1}|\boldsymbol{c}_1^\top ||g(\bm{x})|\bm{1}_{2n}.
\end{aligned}
\label{eq: robust CBF}
\end{equation}

If Eq. \eqref{eq: robust CBF} is satisfied, the safety set $\mathcal{C}$ is forward invariant with probability $1-\epsilon_{\text{con}}$. As in Eq.~\eqref{eq: robust CBF}, for simplicity, we define the confidence bound for the CBF constraint as:
\begin{equation}
E_{\text{con}} = C^{t\mid t-1}|\boldsymbol{c}_1^\top|\bm{1}_{2n}  + C^{t\mid t-1}|\boldsymbol{c}_1^\top ||g(\bm{x})|\bm{1}_{2n}
\label{eq: represent error bound}
\end{equation}
Then we have the CBF with conformal prediction bound~\eqref{eq: CBF with conformal}.

Since the states of the preceding HDV of a CAV can be directly sensed by the CAV, the conformal prediction error primarily impacts the safety of the following HDVs. Using the CBF constraint formulation with conformal prediction bounds as given in Eq.~\eqref{eq: CBF with conformal}, and the Lie derivatives of the CBF candidate calculated above, the CBF constraints for HDVs are given by:
\begin{equation}
\begin{aligned}
&L_{f} h_{i,\text{suf}} + L_{g} h_{i,\text{suf}} \boldsymbol{u} + \alpha_{i,\text{CBF}}(h_{i,\text{suf}}) + \sigma_i \geq E_{\text{con}}, \\
&i \in \Omega_{\mathcal{H}} \setminus \Omega_{\mathcal{P}, i_{\text{first}}},
\end{aligned}
\label{eq:cbf_lcc}
\end{equation}
\noindent where the class $\mathcal{K}_{\infty}$ function $\alpha_{i,\text{CBF}}(\cdot)$ is set to a linear function with a positive coefficient $\gamma_{i,\text{CBF}}$ serving as the parameters to be trained. 
$\sigma_{i}$ is the slack variable for the HDV safety constraint to avoid conflict with CAV safety constraints. 
For the CAV indexed by $j_k$, the vector $\boldsymbol{u}=\left[\hat{u}_{j_1},\cdots,u_{j_k},\cdots,\hat{u}_{j_{m_{\text{ahead}}}}\right]^\top$, with $m_{\text{ahead}}=|\Omega_{\mathcal{F},i}\cap \Omega_{\mathcal{C}}|$, contains the control inputs for the CAVs ahead of HDV $i$. Except for CAV $j_k$, the control inputs of the other CAVs are estimated by the conformal prediction module.
To better tailor the derived CBF constraint to the DRL settings, we rewrite Eq.~\eqref{eq:cbf_lcc} as follows:
\begin{equation}
\begin{aligned}
&L_{f}h_{i,\text{suf}} + L_{g}h_{i,\text{suf}}(\boldsymbol{u}_{\text{safe}} + \boldsymbol{u}_{\text{RL}}) + \gamma_{i,\text{CBF}}h_{i,\text{suf}}+\sigma_i \geq E_{\text{con}},\\
&i\in\Omega_{\mathcal{H}}\setminus\Omega_{\mathcal{P},i_{\text{first}}}
\end{aligned}
\label{eq:cbf_HDV}
\end{equation}

Using the distributed multi-agent DRL-based nominal controller, HDV safety constraints \eqref{eq:cbf_HDV} and CAV safety constraints \eqref{eq:cbf_CAV}, we can formulate the controller that enforces cooperative safety for mixed-autonomy platoon using quadratic programming in the following part.

\subsubsection{QP Problem Formulation}
\label{subsec:QP}
Using the safety constraints, we formulate the following optimization problem to construct the safety layer for CAV indexed in $j_k$ as follows:
\begin{subequations}
\begin{align}
&\min_{\boldsymbol{u}_{\text{safe}}, \sigma_i} \|\boldsymbol{u}_{\text{safe}}\|^2 + \sum_{i\in \Omega_{\mathcal{F},j_k}\cap\Omega_{\mathcal{H}}}b_i\sigma_i^2 \label{eq: optimization objective}\\
\text{s.t.}\quad  & a_{\max}\geq u_{\text{safe},j} + u_{\text{RL},j}\geq a_{\min},\notag\\
& \quad \quad  j\in(\Omega_{\mathcal{F},j_k}\cup\Omega_{\mathcal{P},j_k}\cup \{j_k\})\cap\Omega_{\mathcal{C}}\label{eq: actuator limitation}\\
&v_{j-1} - v_j -\tau  (u_{\text{safe},j} + u_{\text{RL},j}) + \gamma_{j,\text{CAV}}(s_j-\tau u_j) \geq 0,\notag\\
&\quad \quad j\in(\Omega_{\mathcal{F},j_k}\cup\Omega_{\mathcal{P},j_k}\cup \{j_k\})\cap\Omega_{\mathcal{C}} \label{eq: optimization CAV safety}\\
&v_{i-1} - v_i - \tau\hat{\mathbb{F}}_i(s_i,v_i,v_{i-1}) -\sum_{j\in\Omega_{\mathcal{S},i}}k_{i,j,\text{CBF}}\Big(v_{j-1} - v_j\Big)\notag \\
&-\sum_{j\in\Omega_{\mathcal{S},i}}\tau k_{i,j,\text{CBF}}u_j+\gamma_{i,\text{HDV}}h_{i,\text{suf}}+\sigma_i\geq E_{\text{con}},\notag\\
&\quad \quad i \in \Omega_{\mathcal{F},j_k}\cap\Omega_{\mathcal{H}}
\label{eq: optimization HDV safety}
\end{align}
\label{eq: control optimization}
\end{subequations}
where the objective function in Eq. \eqref{eq: optimization objective} is designed to minimize the deviation of the safety control input from the nominal control input and the magnitude of the relaxation coefficients. Additionally, Eq. \eqref{eq: actuator limitation} is the actuator limitation specific to each CAV. Moreover, Eq. \eqref{eq: optimization CAV safety} represents the CAV safety constraints, with the quantity of these constraints matching the number of CAVs within the platoon. Simultaneously, Eq. \eqref{eq: optimization HDV safety} represents the HDV safety constraints, with the count of these constraints corresponding to the number of following HDVs relative to the lead CAV.

\subsubsection{Differentiable QP Layer}
Next, we introduce a differentiable QP that allows the backpropagation of the QP solution from Section~\ref{subsec:QP}, facilitating the training process for MARL. Such treatment can help better integrate the proposed QP and the MARL framework. 

Let the loss function be denoted by $\ell$, and for the CAV indexed by $j_k$, the optimal solution of the CBF-QP as $\boldsymbol{w}_{j_k}^{\star}= \left(\boldsymbol{u}_{\text{safe}}, \sigma_1,\cdots,\sigma_n\right)$. Define  $\bm{\theta}_{j_k}=\left\{\bm{\theta}_{j_k,\text{RL}},\bm{\theta}_{j_k,\text{CBF}}\right\}$ as the trainable parameters of the DRL actor network and the CBF-QP controller, respectively.  
To train these parameters, we aim to compute the following gradients, which are derived using the chain rule:
\begin{align}
    \frac{\partial \ell}{\partial \bm{\theta}_{j_k,\text{RL}}} &= 
    \frac{\partial \ell}{\partial u_{j_k}}\frac{\partial u_{j_k}}{\partial \bm{\theta}_{j_k,\text{RL}}} = \frac{\partial \ell}{\partial u_{j_k}}\Big(\frac{\partial u_{j_k,\text{RL}}}{\partial \bm{\theta}_{j_k,\text{RL}}} + \frac{\partial u_{j_k,\text{safe}}}{\partial \bm{\theta}_{j_k,\text{RL}}} \Big) \notag \\ 
    & = \frac{\partial \ell}{\partial u_{j_k}}\Big(\frac{\partial u_{j_k,\text{RL}}}{\partial \bm{\theta}_{j_k\text{RL}}} + \frac{\partial u_{j_k,\text{safe}}}{\partial u_{j_k,\text{RL}}}\frac{\partial u_{j_k,\text{RL}}}{\partial \bm{\theta}_{j_k,\text{RL}}} \notag
    \\&+ \frac{\partial u_{j_k,\text{safe}}}{\partial \bm{\theta}_{j_k,\text{CBF}}}\frac{\partial \bm{\theta}_{j_k,\text{CBF}}}{\partial \bm{\theta}_{j_k,\text{RL}}}\Big) \notag \\     
    & = \frac{\partial \ell}{\partial u_{j_k}}\Big(\frac{\partial u_{j_k,\text{RL}}}{\partial \bm{\theta}_{j_k,\text{RL}}} + \frac{\partial u_{j_k,\text{safe}}}{\partial u_{j_k,\text{RL}}}\frac{\partial u_{j_k,\text{RL}}}{\partial \bm{\theta}_{j_k,\text{RL}}}\Big) , \label{eq:partial_RL}\\
    \frac{\partial \ell}{\partial \bm{\theta}_{j_k,\text{CBF}}} &= 
    \frac{\partial \ell}{\partial u_{j_k}}\frac{\partial u_{j_k}}{\partial \bm{\theta}_{j_k,\text{CBF}}} = \frac{\partial \ell}{\partial u_{j_k}}\left(\frac{\partial u_{j_k,\text{RL}}}{\partial \bm{\theta}_{j_k,\text{CBF}}} + \frac{\partial u_{j_k,\text{safe}}}{\partial \bm{\theta}_{j_k,\text{CBF}}}\right) \notag\\
    & = \frac{\partial \ell}{\partial u_{j_k}} \frac{\partial u_{j_k,\text{safe}}}{\partial \bm{\theta}_{j_k,\text{CBF}}} \label{eq:partial_CBF}
\end{align}
where the first equality in both Eq. (\ref{eq:partial_RL}) and Eq. (\ref{eq:partial_CBF}) arises from the application of the chain rule, while the second equality in both equations is based on the relationship $u_{j_k} = u_{j_k,\text{RL}} + u_{j_k,\text{safe}}$. The third equality in Eq. (\ref{eq:partial_RL}) follows Eq. (\ref{eq: control optimization}), where $u_{j_k,\text{safe}}$ is treated as a function of $u_{j_k,\text{RL}}$ and $\bm{\theta}_{j_k,\text{CBF}}$. The final equality in Eq. (\ref{eq:partial_RL}) holds because $\frac{\partial \bm{\theta}_{j_k,\text{CBF}}}{\partial \bm{\theta}_{j_k,\text{RL}}} = 0$, as the CBF-QP parameters $\bm{\theta}_{j_k,\text{CBF}}$ are independent of the actor network.
It is important to point out that $\frac{\partial \ell}{\partial u_{j_k}}$ can be easily derived from the loss function, and $\frac{\partial u_{j_k,\text{RL}}}{\partial \bm{\theta}_{j_k}}$ is determined by the architecture of the actor network. Thus, the focus is on calculating $\frac{\partial u_{j_k,\text{safe}}}{\partial u_{j_k,\text{RL}}}$ and $\frac{\partial u_{j_k,\text{safe}}}{\partial \bm{\theta}_{j_k,\text{CBF}}}$, which are components of $\frac{\partial w_{j_k}^{\star}}{\partial u_{j_k,\text{RL}}}$  and $\frac{\partial w_{j_k}^{\star}}{\partial \bm{\theta}_{j_k,\text{CBF}}}$ as discussed next.

To compute these gradients, we reformulate the CBF-QP from Eq.~\eqref{eq: control optimization} in a more general form, with the decision variable denoted as  $\boldsymbol{w}_{j_k}=\left(\boldsymbol{u}_{\text{safe}}, \sigma_1,\cdots,\sigma_n\right)$: 
\begin{equation}
    \begin{aligned}
    \min_{\boldsymbol{w}_{j_k}} \quad & \frac{1}{2} \boldsymbol{w}_{j_k}^T Q \boldsymbol{w}_{j_k} \\
    \text { subject to } \quad & G \boldsymbol{w}_{j_k} \leq q(u_{j_k,\text{RL}}, \bm{\theta}_{j_k,\text{CBF}})
    \end{aligned}
    \label{safe QL}
\end{equation}
where $Q, G, q$ are the parameters of the QP formulation, $\bm{\theta}_{j_k,\text{CBF}} = [\gamma_{j,\text{CAV}},\gamma_{i,\text{HDV}}],j\in(\Omega_{\mathcal{F},j_k}\cup\Omega_{\mathcal{P},j_k}\cup \{j_k\})\cap\Omega_{\mathcal{C}},i \in \Omega_{\mathcal{F},j_k}\cap\Omega_{\mathcal{H}}$ represents the parameters of the CBF-based safety constraints. Here, only $q$ is a function of $u_{j_k,\text{RL}}$ and $\bm{\theta}_{j_k,\text{CBF}}$. Following the approach in~\cite{xiao2023barriernet}, the calculation of $\frac{\partial \boldsymbol{w}_{j_k}^{\star}}{\partial u_{j_k,\text{RL}}}$ and $\frac{\partial \boldsymbol{w}_{j_k}^{\star}}{\partial \bm{\theta}_{j_k,\text{CBF}}}$ can be achieved by differentiating the KKT conditions with respect to equality, i.e.,
\begin{equation}
\begin{aligned}
Q w_{j_k}^{\star}+G^T \lambda^{\star} & =0, \\
\mathbb{D}\left(\lambda^{\star}\right)\left(G \boldsymbol{w}_{j_k}^{\star}-q(u_{j_k,\text{RL}}, \bm{\theta}_{j_k,\text{CBF}})\right) & =0,
\end{aligned} \label{eq:KKT}
\end{equation}
where $\boldsymbol{w}_{j_k}^{\star}$ and $\lambda^{\star}$ denote the optimal primal and dual variables, and $\mathbb{D}(\lambda^\star)$ is a diagonal matrix derived from vector $\lambda^\star$. 

Since the process for calculating $\frac{\partial u_{j_k,\text{safe}}}{\partial u_{j_k,\text{RL}}}$ and $\frac{\partial u_{j_k,\text{safe}}}{\partial \bm{\theta}_{j_k,\text{CBF}}}$ is similar, we will only focus on illustrating the calculation of $\frac{\partial u_{j_k,\text{safe}}}{\partial u_{j_k,\text{RL}}}$. To compute $\frac{\partial \boldsymbol{w}_{j_k}^{\star}}{\partial \bm{\theta}_{j_k,\text{CBF}}}$, we simply replace $u_{j_k,\text{RL}}$ with $\bm{\theta}_{j_k,\text{CBF}}$ in Eq. \eqref{derivative of KKT}. 
Next, we take the derivative of Eq. \eqref{eq:KKT} with respect to the RL action $u_{j_k,\text{RL}}$ and express it in the matrix form as follows:
\begin{equation}
\begin{aligned}
&  {\left[\begin{array}{l}
\displaystyle \frac{\partial \boldsymbol{w}_{j_k}^{\star}}{\partial u_{j_k,\text{RL}}} \\
 \displaystyle \frac{\partial  \lambda^{\star}}{\partial u_{j_k,\text{RL}}} 
\end{array}\right]=}  K^{-1}\left[\begin{array}{c}
\mathbb{O} \\
\mathbb{D}\left(\lambda^{\star}\right)\displaystyle \frac{\partial  q(u_{j_k,\text{RL}},\bm{\theta}_{j_k,\text{CBF}})}{\partial u_{j_k,\text{RL}}} 
\end{array}\right],
\end{aligned}
\label{derivative of KKT}
\end{equation}
with $K = \left[\begin{array}{cc}
Q & G^T \\
\mathbb{D}\left(\lambda^{\star}\right) G & \mathbb{D}\left(G \boldsymbol{w}_{j_k}^\star-q(u_{j_k,\text{RL}},\bm{\theta}_{j_k,\text{CBF}})\right)  
\end{array}\right]$. 

By leveraging the partial derivatives of the QP layer, the parameters of both the DRL and CBF constraints can be updated simultaneously, enabling adaptive adjustments to the training environment.

\section{Simulation and Result Analysis}
\label{sec: Simulation Results}
In this section, simulation experiments are conducted to evaluate the performance of our proposed safe MARL-based controller. Specifically, we are interested in the tradeoff between safety in safety-critical scenarios and efficiency in normal operating scenarios. The simulation setup is illustrated in Fig. \ref{fig:CAV_overview}, where CAVs are indexed as $\Omega_{\mathcal{C}}=\left\{2,4\right\}$ and HDVs are indexed as $\Omega_{\mathcal{H}}=\left\{1,3,5,6,7\right\}$, with the head vehicle indexed as $0$. 
Since system-level safety has been rarely considered in existing literature \cite{zhou2024enhancing}, we evaluate the performance of our proposed controller by comparing the following methods:
\begin{enumerate}
    \item[M1.] Pure car-following model, where CAVs adopt the same car-following behavior as HDVs.
    \item[M2.] MARL w/o safety guarantees, where the controllers for CAVs are based on MAPPO without integrating with the safety layer.
    \item[M3.] Safe-MARL non-cooperative, where CAV controllers employ MAPPO with non-cooperative CBFs as described in \cite{zhou2024enhancing}.
    \item[M4.] Safe-MARL w/o CP: CAV controllers using MAPPO with cooperative CBFs but without conformal prediction.
    \item[M5.] Safe-MARL with CP: CAV controllers using MAPPO with cooperative CBFs and conformal prediction. 
\end{enumerate}

Section \ref{subsec: Training Setting} introduces the simulation settings. Section \ref{subsec: Performance of Safety Enhancement} demonstrates the safety enhancement performance in safety-critical scenarios. Section~\ref{subsec: Analysis of Safety-Utility Trade-off} analyzes the impact of the controller on control efficiency. 

\subsection{Simulation Settings}
\label{subsec: Training Setting}

\noindent\emph{Environment Settings.} 
Without loss of generality, we choose the Full Velocity Difference
Model (FVD) \cite{jiang2001full} as the car-following model $\mathbb{F}$, following the setting of LCC~\cite{wang2021leading,zhou2024enhancing}. Note that the FVD can be replaced by any car-following model without influencing the applicability of the proposed method. The FVD reads as follows:
\begin{equation}
    \mathbb{F}(\cdot)=\alpha\left(V\left(s_i(t)\right)-v_i(t)\right)+\beta (v_{i-1}(t)-v_{i}(t)),
\end{equation}
where constants $\alpha,\beta>0$ represent car-following gains.
$V(s)$ denotes the spacing-dependent desired velocity of HDVs with the form given in Eq. (\ref{eq:optimal_speed}), where $s_{\mathrm{st}}$ and $s_{\mathrm{go}}$ represent the spacing thresholds for stopped and free-flow states, respectively, and  $v_{\mathrm{max}}$ denotes the free-flow velocity. 
\begin{equation}
V(s)= 
\begin{cases}
0, & s \leq s_{\mathrm{st}} \\
\displaystyle\frac{v_{\max }}{2}\left(1-\cos \left(\pi \frac{s-s_{\mathrm{st}}}{s_{\mathrm{go}}-s_{\mathrm{st}}}\right)\right), & s_{\mathrm{st}}<s<s_{\mathrm{go}} \\ 
v_{\mathrm{max}}, & s \geq s_{\mathrm{go}}
\end{cases} \label{eq:optimal_speed}
\end{equation}
The parameters of FVD is set as $\alpha = 0.6,\beta=0.9,s_{\text{st}}=5,s_{\text{go}}=35$. The equilibrium spacing and velocity are $20$ m and $15$ m/s.

\begin{figure}[t]
    \centering
    \includegraphics[width=8cm]{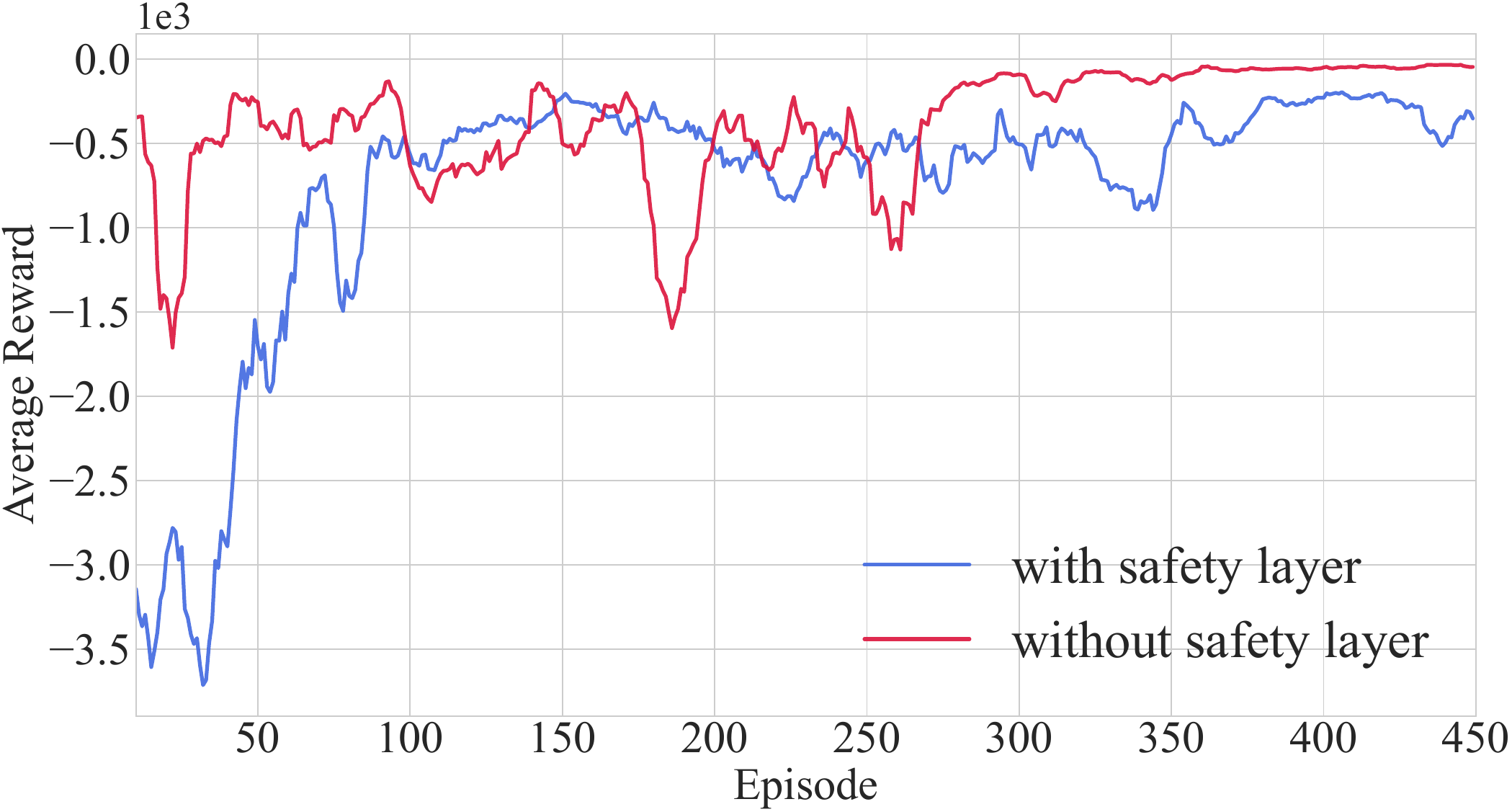}
    \caption{Accumulated training rewards per episode.}
    \label{fig:training}
\end{figure}

\noindent\emph{MARL Training Settings}.  In the training scenario, we utilize a random velocity disturbance setup, wherein the head vehicle's velocity disturbance per time step is sampled from a Gaussian distribution with zero mean and a standard deviation of $0.2$\,m/s independently at each time step within an episode. 
The learning rates for both the actor and critic networks are initially set at 0.0003, with a linear decay schedule applied to facilitate training. This approach ensures that the performance of the algorithms is not highly sensitive to the initial learning rate, allowing us to select rates that expedite convergence. The configuration includes two agents with training batch size $2048$. The decay factor for the advantage function is set at $0.95$, and the PPO clipping parameter is fixed at $0.2$. The experience replay buffer size is set at $10$. The training consists of $450$ episodes, each comprising $1000$ steps with a step duration of $0.1$ seconds. The reward weighting coefficients in Eq.~\ref{eq: global reward} and Eq.~\ref{eq: local reward} are set as follows: $w_{\text{global}} = 0.1$, $w_{\text{local}} = 0.9$, $w_{\text{efficacy}}=1$, and $w_{\text{safety}}=1$.  The accumulated training rewards are displayed in Fig. \ref{fig:training}, demonstrating that both the MARL algorithms, with and without safety guarantees, converge after 450 episodes of training.

\noindent \emph{Robust QP Parameters.} The actuation bounds are set to $a_{\min} = -5$ and $a_{\max}= 5$. The CBF parameters are set to $\tau = 0.3$ and $k_{i,\text{CBF}}=0.4, i\in \left\{2,3,4,5,6\right\}$. 

\begin{figure}[t]
    \centering\includegraphics[width = 8cm]{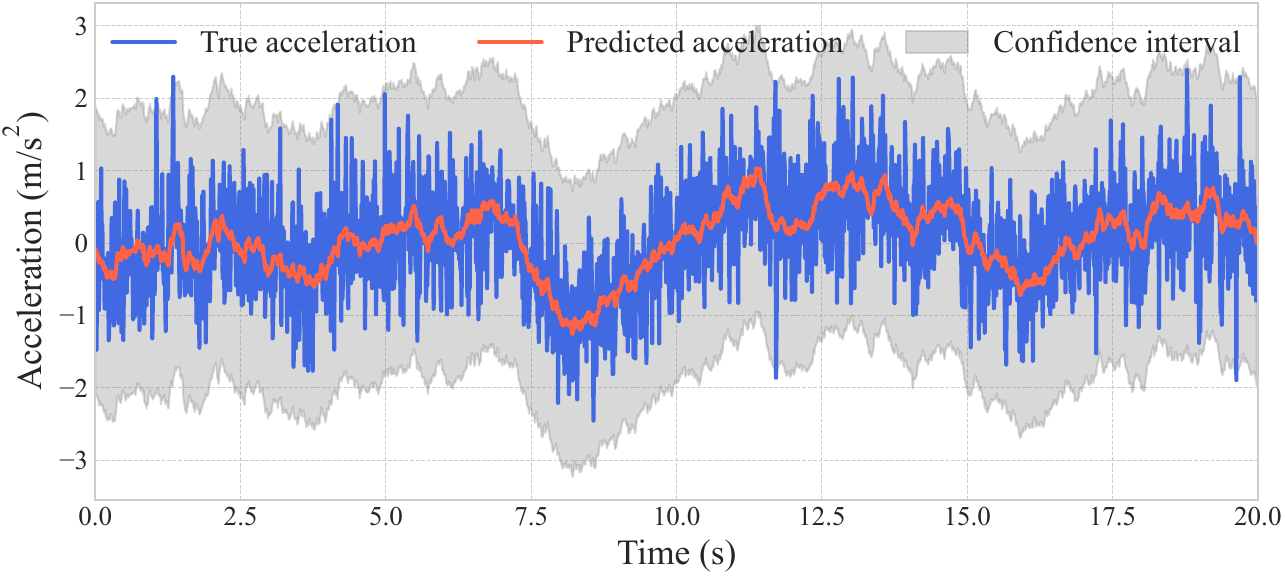}
    \caption{Conformal prediction results for the predicted acceleration with conformal prediction bounds for HDV $1$.}
    \label{fig:CP}
\end{figure}

\noindent \emph{Conformal Prediction Configurations.} 
We employ a fully connected neural networks to predict the behaviors of both HDVs and CAVs. The network utilizes the Adam optimizer with a learning rate of $0.001$. The input features for the neural network include the acceleration, velocity, and spacing of the vehicle being predicted, as well as the velocity of the preceding vehicle. We set the failure probability, $\epsilon_{\text{con}}$, as $0.01$. Fig. \ref{fig:CP} shows the conformal prediction performance on a testing dataset for HDV $1$. The results indicate that the conformal prediction effectively tracks the trend of HDV behaviors, and the conformal prediction bounds accurately contain the prediction errors.

\subsection{Performance of Safety Enhancement}
\label{subsec: Performance of Safety Enhancement}
In this subsection, the trained cooperative safe MARL agent is evaluated in two safety-critical LCC scenarios: 
\begin{itemize} 
    \item \textbf{Scenario 1}: This scenario addresses the situation where the preceding HDV of CAV $1$ may brake urgently to avoid collisions with an unexpected cutting-in vehicle or crossing pedestrians. This abrupt braking could reduce the distance to the following CAV to less than the safety distance.
    \item \textbf{Scenario 2}: This scenario describes the situation where the following HDV (indexed by $5$) of CAV $2$ might suddenly accelerate due to human errors caused by driving fatigue, potentially leading to safety issues with the vehicle ahead.
\end{itemize}
Note that these two scenarios are essentially different from the scenarios for training. 

To evaluate the performance of the proposed method, we conduct two types of experiments within each safety-critical scenario: (i) the analysis of safety-guarantee regions, i.e., the region of the acceleration and duration of the disturbances such that the LCC system is safe under a control policy, 
and (ii) specific case studies to compare the safety performance of different benchmarks in typical safety-critical scenarios. 
Our results are presented as follows. 

\subsubsection{Safety-Guaranteed Regions}
\begin{figure}[htp!]
    \centering
    \subcaptionbox{Safe region for Scenario 1}{
            \includegraphics[width=7cm]{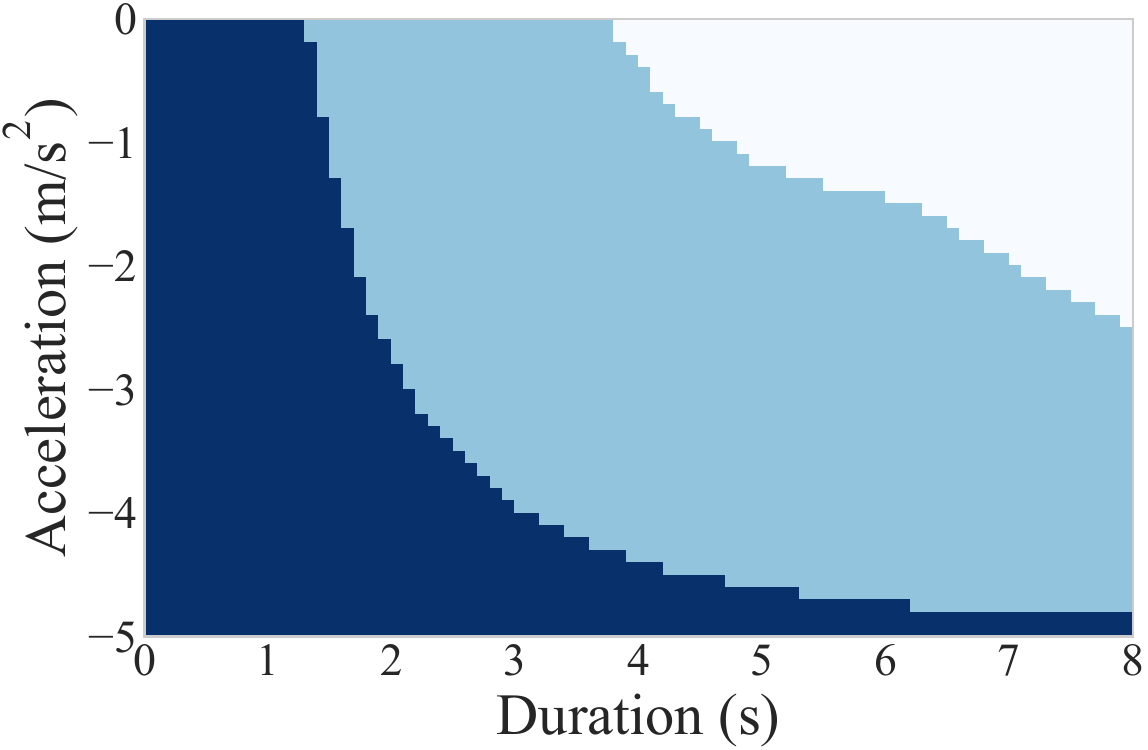}
        }
    \subcaptionbox{Safe region for Scenario 2}{
            \includegraphics[width=7cm]{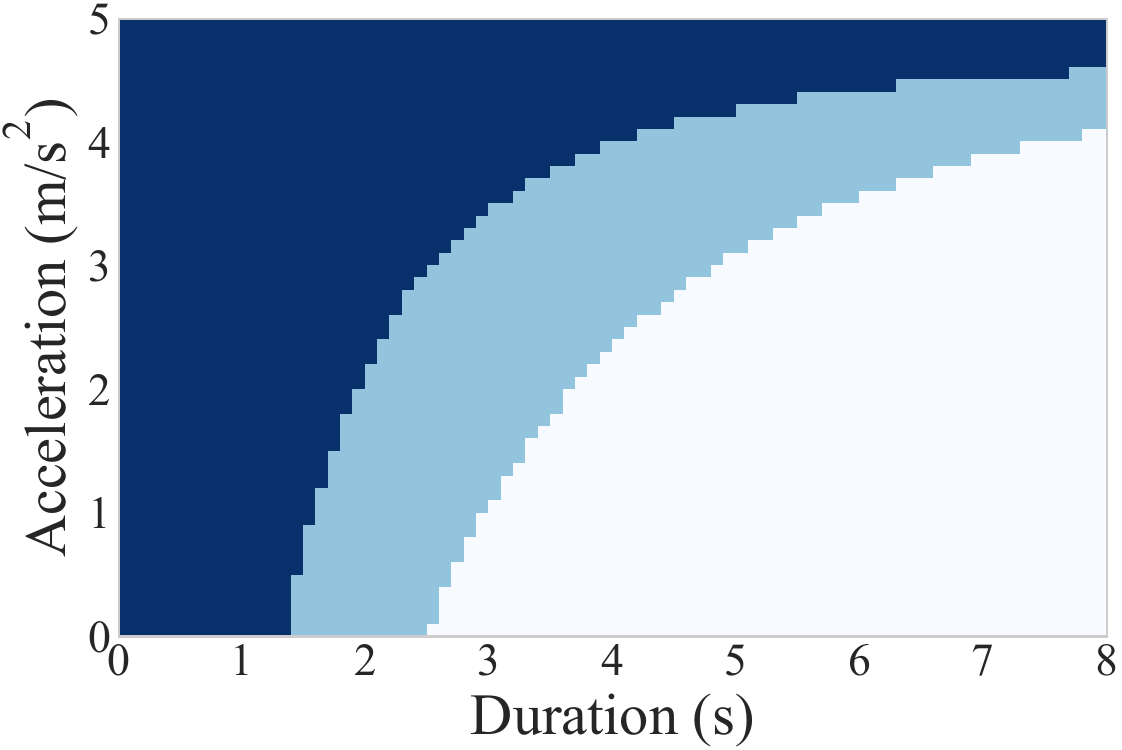}
        }
    \caption{Safety-guaranteed regions associated with two safety-critical scenarios.
    The dark blue region represents the safety region for the CAVs equipped with MARL controllers but without the safety layer (M2). The light blue region indicates the enhanced safety regions achieved through our proposed method (M5). The white region denotes areas that are considered unsafe.}
    \label{fig:safety region}
\end{figure}

To demonstrate the effectiveness of the proposed cooperative CBF design, we compare the safety regions with and without the safety layer (i.e., benchmarks M5 and M2, respectively) in both safety-critical scenarios. Fig. \ref{fig:safety region} (a) illustrates the performance in Scenario 1, where the preceding vehicle of CAV $1$ brakes. The figure shows that the incorporation of the proposed safety layer leads to an expansion of the safety region by nearly 100\%. Fig. \ref{fig:safety region} (b) demonstrates the performance in Scenario 2, whereby HDV $5$ performs irrational acceleration. Specifically, the safety region expands by nearly 70\%. The significant expansion of the safety region occurs because the two CAVs are positioned ahead of HDV 5 and can collaborate to enhance safety more effectively. Overall, our simulation shows that the incorporation of the safety layer can enable the mixed-autonomy platoon to safely handle a much wider range of safety-critical situations. 

\begin{figure}[htp]
    \centering
    \subcaptionbox{Safety-critical scenario 1}{
            \includegraphics[width=6cm]{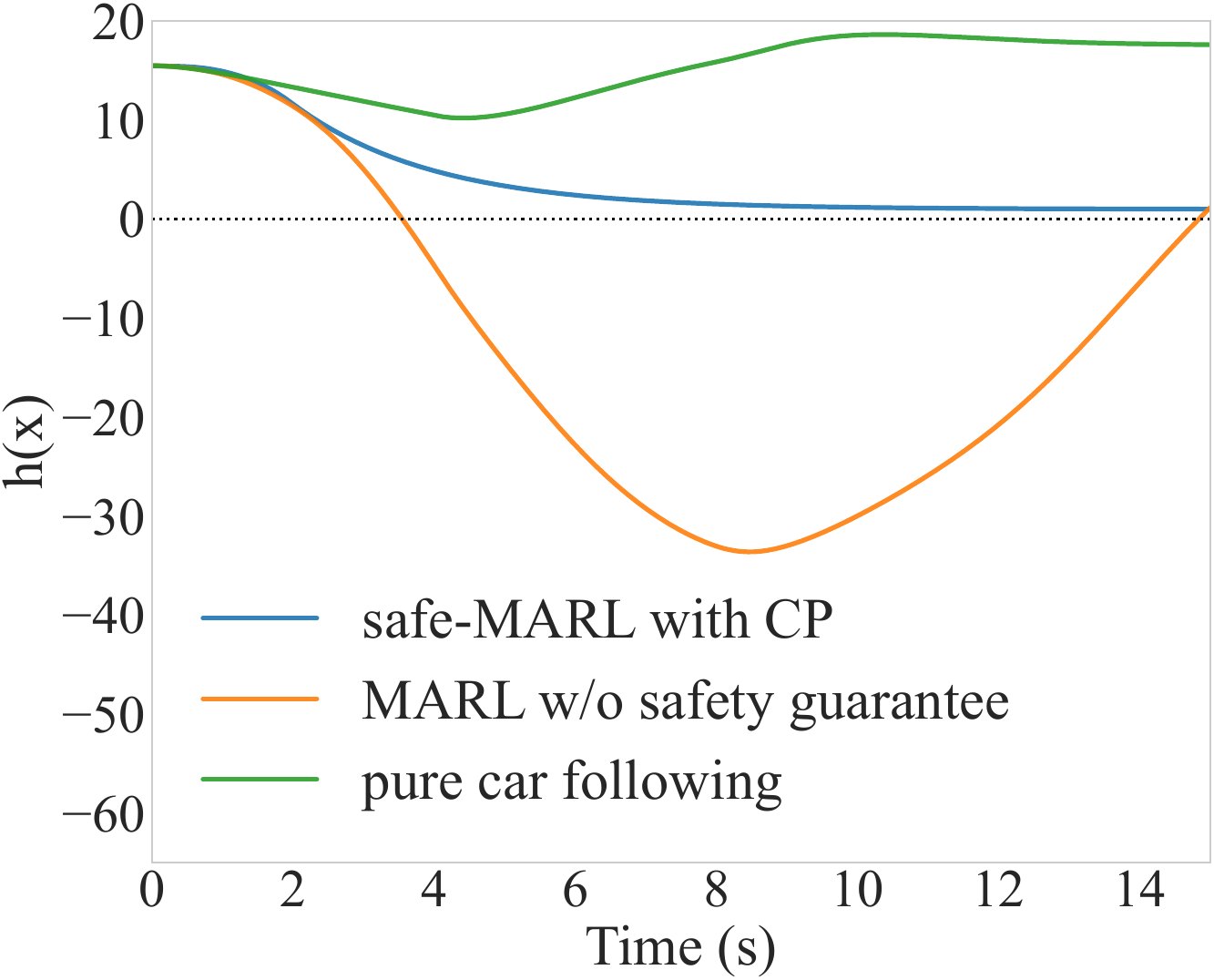}
        }
    \subcaptionbox{Safety-critical scenario 2}{
            \includegraphics[width=6cm]{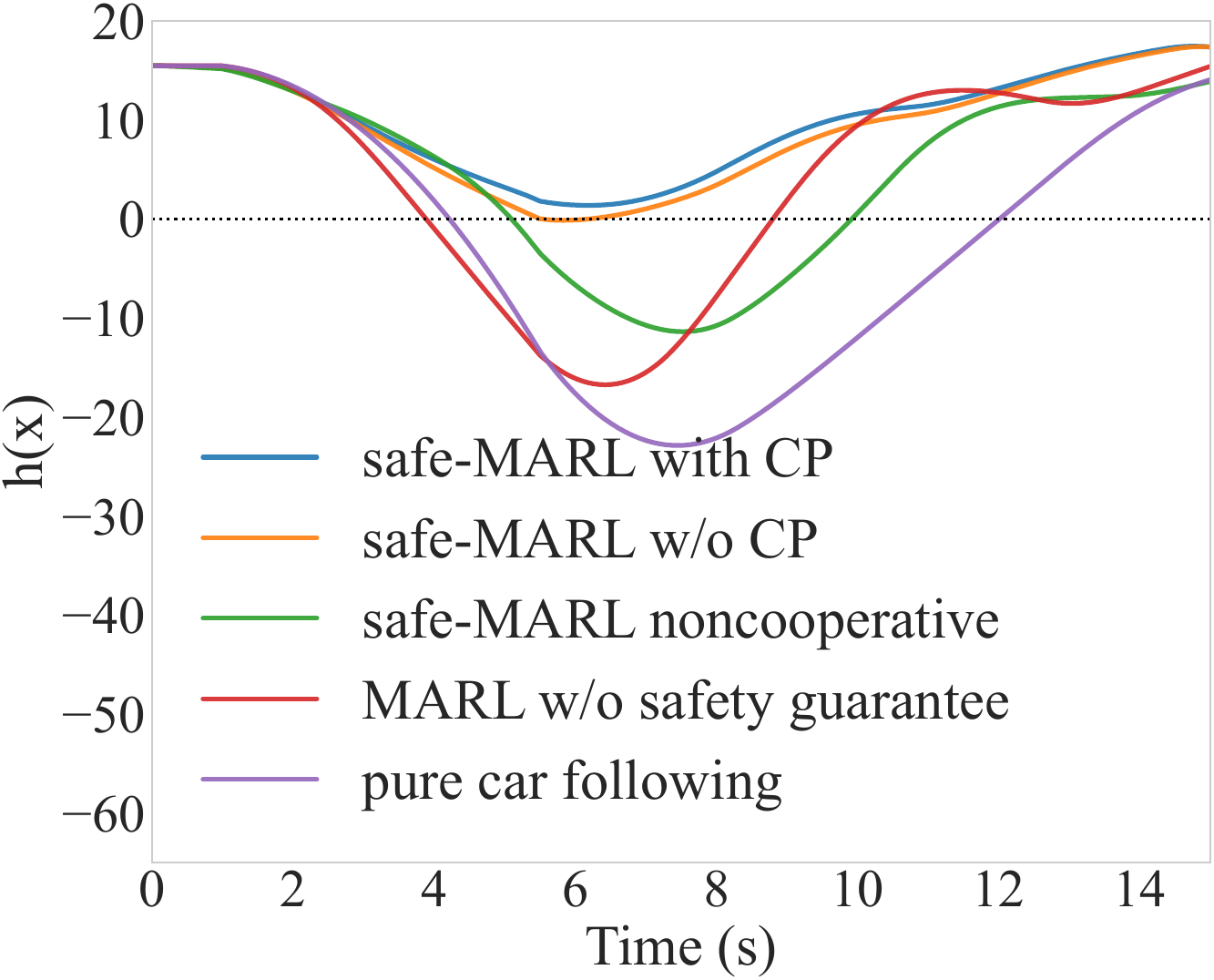}
        }
    \caption{Values of the CBF candidates in the two safety-critical scenarios.}
    \label{fig:hx}
\end{figure}

\begin{figure}[htp]
    \centering
    \subcaptionbox{Safety-critical scenario 1}{
            \includegraphics[width=6cm]{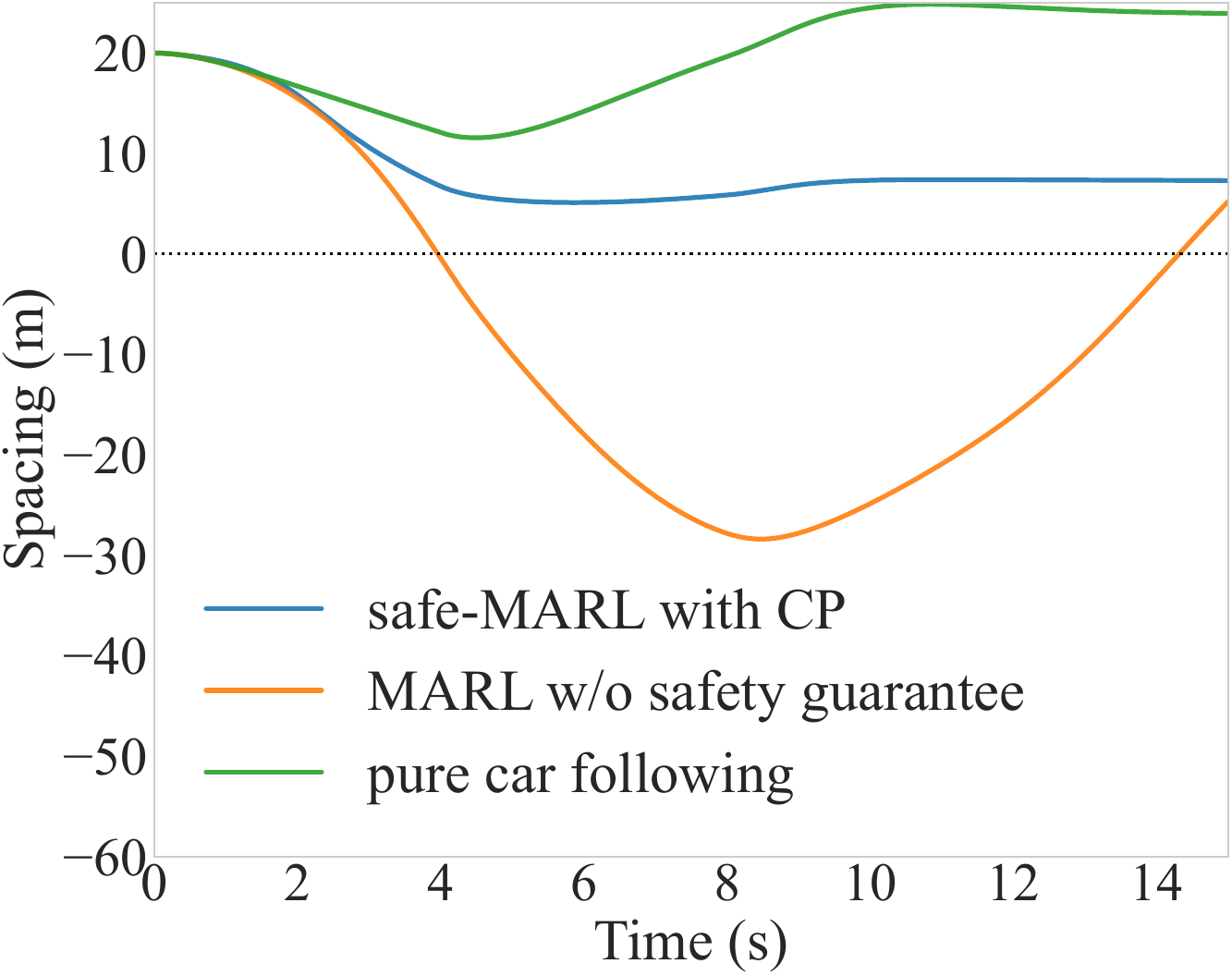}
        }
    \subcaptionbox{Safety-critical scenario 2}{\includegraphics[width=6cm]{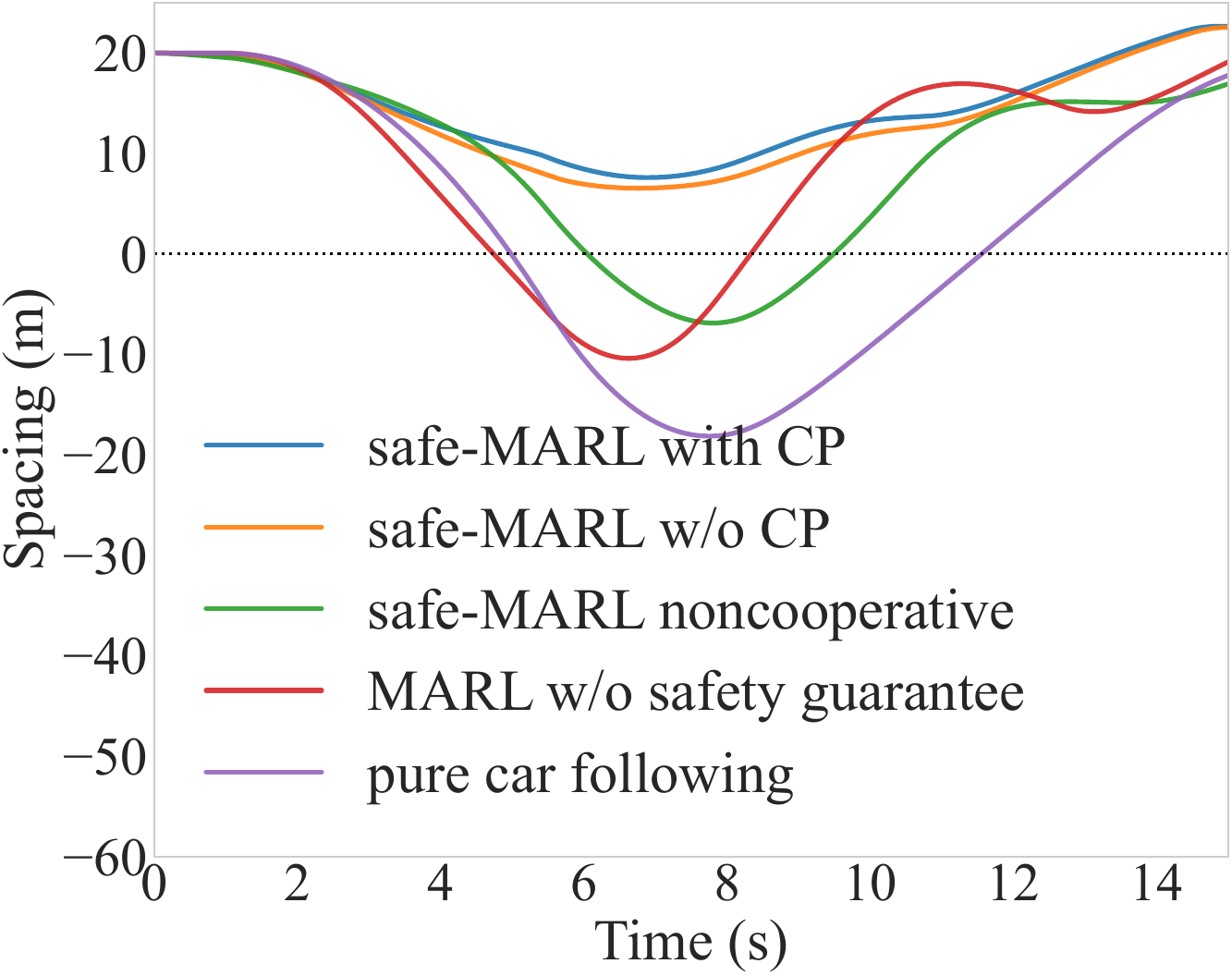}
        }
    \caption{Values of the spacings in the two safety-critical scenarios.}
    \label{fig:spacing}
\end{figure}

\begin{figure}[htp]
    \centering
    \subcaptionbox{Safe MARL with cooperation (M5)}{
            \includegraphics[width=6cm]{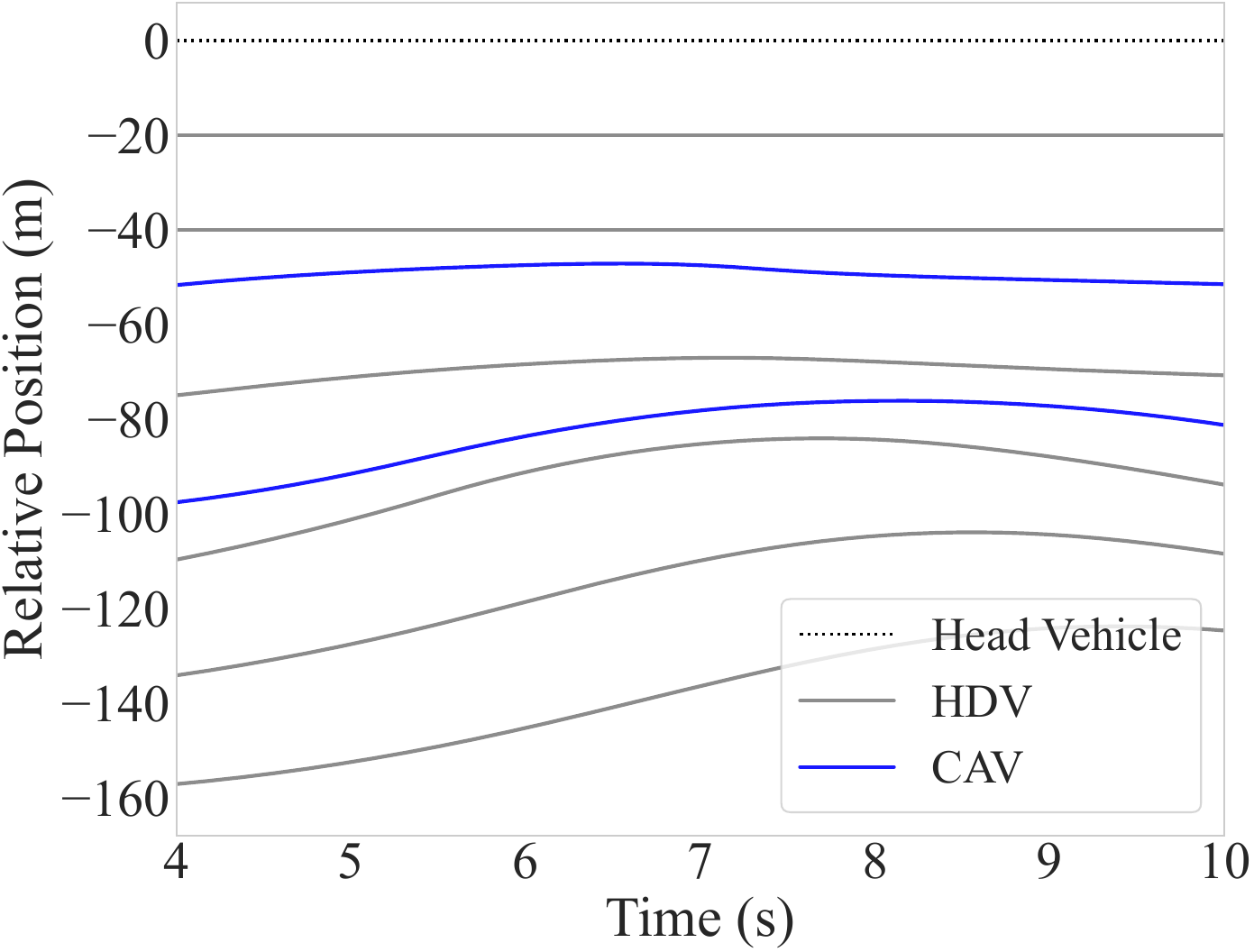}
        }
    \subcaptionbox{Safe MARL without cooperation (M3)}{
            \includegraphics[width=6cm]{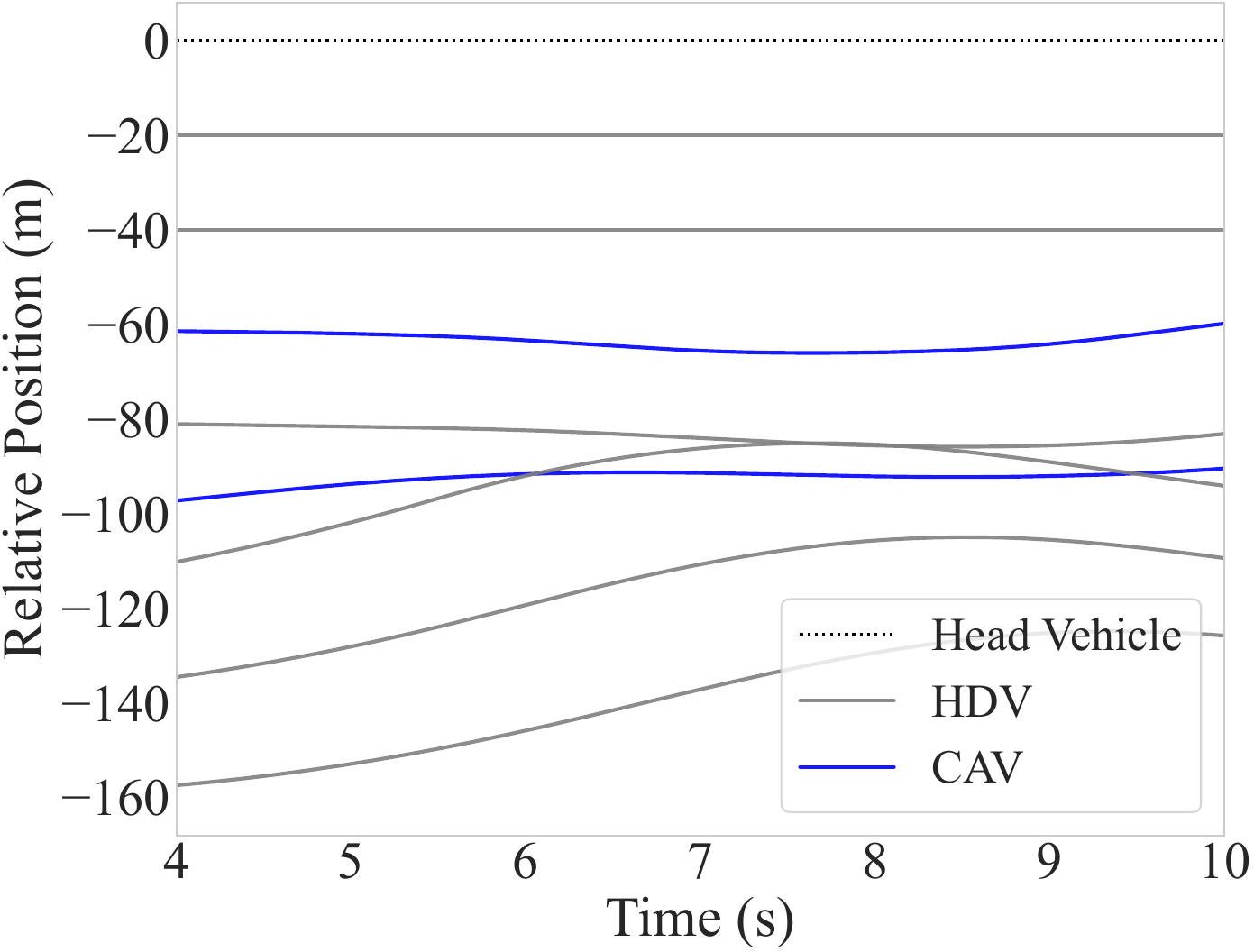}
        }
    \caption{Relative position of all vehicles in the mixed-autonomy platoon with respect to the head vehicle, depicting the safety-critical scenario 2 where the HDV following CAV 2 accelerates due to human error. 
    }
    \label{fig:all_vehicle_spacing}
\end{figure}

\subsubsection{Case Studies for Scenario 1}  
In Scenario 1, the preceding vehicle of CAV $1$, i.e., HDV $0$, brakes urgently with a deceleration of $-3\text{m/s}^2$ for $4$s, followed by an acceleration back to its equilibrium velocity over another~$4$s.  
In this scenario, we demonstrate the value of using CBF for safety guarantees by comparing three control strategies: safe MARL (M5), MARL without safety guarantee (M2), and pure car-following platoon (M1). Note that we do not evaluate the other two benchmarks here since the value of cooperative safety is not significant in this scenario, as the braking of HDV $0$ mainly influences CAV $1$. The results are illustrated in Fig. \ref{fig:hx} (a) and Fig. \ref{fig:spacing} (a).


It can be seen that MARL without a safety layer fails to prevent collisions in extreme cases, resulting in negative spacing between vehicles. The pure car-following setting, in contrast, maintains safety and then returns to the equilibrium spacing. With our proposed method, not only are the CBF candidate values and spacing greater than $0$, but the headway is also reduced, enhancing traffic efficiency.

\subsubsection{Case Studies for Scenario 2}
In this scenario, HDV $5$ performs irrational acceleration with a rate of $2.5\text{m/s}^2$ starting at $1$s for a duration of $4.5$s. 
Fig. \ref{fig:hx} (b) and \ref{fig:spacing} (b) depict the resulting values of the CBF candidates and the spacing from all benchmarks. 
We can see from these two figures that safe-MARL approaches with (M5) and without CP (M4) can ensure safety and forward invariance by ensuring positive spacing and positive CBF candidate values over time, and safety can be slightly improved by incorporating CP. 
However, both the pure car-following setting (M1) and MARL without a safety layer (M2) fail to prevent collisions, as indicated by the negative spacing between the HDV and its preceding vehicle. This highlights the value of the safety layer. 

Moreover, the safe-MARL approach with non-cooperative CBF (M3 \cite{zhou2024enhancing}) can avoid collision but yield a worse safety guarantee by rendering the CBF candidate negative. This shows that the lack of cooperation between the two CAVs weakens the safety guarantee.  
To better illustrate the value of cooperative safety, we demonstrate the evolution of the relative positions of all vehicles resulting from the proposed safe-MARL approach (M5) and the safe-MARL approach with non-cooperative CBF (M3), respectively, in Fig.~\ref{fig:all_vehicle_spacing} (a) and Fig.~\ref{fig:all_vehicle_spacing} (b). 
We can see that when HDV $5$ following CAV $2$, both preceding CAVs accelerate cooperatively to avoid a collision. However, in Fig.~\ref{fig:all_vehicle_spacing} (b), CAV 1 fails to respond to the safety-critical scenario, resulting in a collision. These results further highlight the value of cooperation of CAVs in enhancing the safety of mixed-autonomy platoons.

\subsection{Analysis of Safety-Utility Trade-off}
\label{subsec: Analysis of Safety-Utility Trade-off}

As indicated in Section \ref{subsec: Performance of Safety Enhancement}, our proposed method enhances \emph{system-level} safety in safety-critical scenarios, which may, nevertheless, negatively impact the control performance in general driving conditions. 
In this subsection, we evaluate the safe MARL-based controller in general driving conditions characterized by a sine-like disturbance scenario, where the head vehicle (HDV $0$) follows a sine-like acceleration pattern with an amplitude of $2$ $\text{m/s}^2$. We use the average time headway for CAVs \cite{fairclough1997effect} and the Average Absolute Velocity Error (AAVE) to show the impact of the safety layer on platoon control efficacy. The average time headway ($s$) is calculated as the average of $\frac{s_i(t)}{v_i(t)}$ over the simulation period for all CAVs. A lower average time headway indicates more efficient traffic flow. The AAVE is used to quantify velocity errors, which is obtained by computing the average of $\left|v_i(t)-v_0(t)\right|$ across all vehicles and the entire simulation period.

Table~\ref{tab: Average Time Headway and Average Absolute Velocity Error (AAVE)} displays the average time headway and AAVE  under the sine-like disturbance scenario. From Table~\ref{tab: Average Time Headway and Average Absolute Velocity Error (AAVE)}, we can see that the resulting average time headway from our proposed controller (M5) is similar to that of MARL without safety guarantee (M2, within a difference of $0.12$s) and is lower than the other benchmarks (M3-M5). Moreover, the resulting AAVE from our proposed controller slightly degrades compared to other benchmarks (M1-M4, within a difference of $0.7$m/s).  This shows that the proposed safety layer has minimal impact on the control performance and achieves satisfactory safety-utility trade-offs.


\begin{table}[ht]
\centering
\caption{Average Time Headway and Average Absolute Velocity Error (AAVE)}
\label{tab: Average Time Headway and Average Absolute Velocity Error (AAVE)}
\begin{tabular}{ccc}
\toprule
            & Average Time Headway (s) & AAVE (m/s) \\ \midrule
Pure car-following   &      2.29       &  3.36 \\ 
MARL w/o safety guarantee &   1.98       & 3.17 \\ 
Safe-MARL non-cooperative &     2.34     & 3.48 \\ 
Safe-MARL w/o CP &     2.21    & 3.69 \\ 
Safe-MARL with CP &    2.10     & 3.83 \\
\bottomrule
\end{tabular}
\end{table}

\section{Conclusion and Future Work}
\label{sec: Conclusion}
In this paper, we propose a cooperative safe MARL-based control strategy for multi-CAVs in mixed-autonomy platoons. First, we design a cooperative CBF that allows CAVs to generate safe actions cooperatively, thereby enhancing system-level safety. Then, we design a conformal prediction module to quantify the states' error disturbance. 
We then integrate the cooperative CBF and the conformal prediction bounds into a MARL framework via a differentiable safety layer. Our proposed method not only improves the cooperative safety for MARL but also simultaneously enhances traffic efficiency.

This research opens several interesting directions for future work. First, each scenario involving different positions and numbers of CAVs in the platoon within the MARL framework requires re-training with a new model, making the current method difficult to implement in real traffic scenarios. Therefore, we are interested in using meta-RL to enhance the generality of the current method. Second, we would like to integrate adaptive control barrier functions \cite{xiao2021adaptive} into the current framework to accommodate time-varying control bounds, thereby enhancing the generality across different scenarios.

\bibliographystyle{IEEEtran}
\bibliography{ref.bib}

\begin{thebibliography}{10}
\providecommand{\url}[1]{#1}
\csname url@samestyle\endcsname
\providecommand{\newblock}{\relax}
\providecommand{\bibinfo}[2]{#2}
\providecommand{\BIBentrySTDinterwordspacing}{\spaceskip=0pt\relax}
\providecommand{\BIBentryALTinterwordstretchfactor}{4}
\providecommand{\BIBentryALTinterwordspacing}{\spaceskip=\fontdimen2\font plus
\BIBentryALTinterwordstretchfactor\fontdimen3\font minus \fontdimen4\font\relax}
\providecommand{\BIBforeignlanguage}[2]{{%
\expandafter\ifx\csname l@#1\endcsname\relax
\typeout{** WARNING: IEEEtran.bst: No hyphenation pattern has been}%
\typeout{** loaded for the language `#1'. Using the pattern for}%
\typeout{** the default language instead.}%
\else
\language=\csname l@#1\endcsname
\fi
#2}}
\providecommand{\BIBdecl}{\relax}
\BIBdecl

\bibitem{deng2023cooperative}
Z.~Deng, K.~Yang, W.~Shen, and Y.~Shi, ``Cooperative platoon formation of connected and autonomous vehicles: Toward efficient merging coordination at unsignalized intersections,'' \emph{IEEE Transactions on Intelligent Transportation Systems}, 2023.

\bibitem{liang2022mas}
J.~Liang, Y.~Li, G.~Yin, L.~Xu, Y.~Lu, J.~Feng, T.~Shen, and G.~Cai, ``A mas-based hierarchical architecture for the cooperation control of connected and automated vehicles,'' \emph{IEEE Transactions on Vehicular Technology}, vol.~72, no.~2, pp. 1559--1573, 2022.

\bibitem{WANG2024104743}
Q.~Wang and K.~Yang, ``Privacy-preserving data fusion for traffic state estimation: A vertical federated learning approach,'' \emph{Transportation Research Part C: Emerging Technologies}, p. 104743, 2024.

\bibitem{tan2024connected}
C.~Tan, Y.~Cao, X.~Ban, and K.~Tang, ``Connected vehicle data-driven fixed-time traffic signal control considering cyclic time-dependent vehicle arrivals based on cumulative flow diagram,'' \emph{IEEE Transactions on Intelligent Transportation Systems}, 2024.

\bibitem{zhao2018platoon}
W.~Zhao, D.~Ngoduy, S.~Shepherd, R.~Liu, and M.~Papageorgiou, ``A platoon based cooperative eco-driving model for mixed automated and human-driven vehicles at a signalised intersection,'' \emph{Transportation Research Part C: Emerging Technologies}, vol.~95, pp. 802--821, 2018.

\bibitem{yang2022eco}
J.~Yang, D.~Zhao, J.~Lan, S.~Xue, W.~Zhao, D.~Tian, Q.~Zhou, and K.~Song, ``Eco-driving of general mixed platoons with cavs and hdvs,'' \emph{IEEE Transactions on Intelligent Vehicles}, vol.~8, no.~2, pp. 1190--1203, 2022.

\bibitem{wang2021leading}
J.~Wang, Y.~Zheng, C.~Chen, Q.~Xu, and K.~Li, ``Leading cruise control in mixed traffic flow: System modeling, controllability, and string stability,'' \emph{IEEE Transactions on Intelligent Transportation Systems}, vol.~23, no.~8, pp. 12\,861--12\,876, 2021.

\bibitem{shi2021connected}
H.~Shi, Y.~Zhou, K.~Wu, X.~Wang, Y.~Lin, and B.~Ran, ``Connected automated vehicle cooperative control with a deep reinforcement learning approach in a mixed traffic environment,'' \emph{Transportation Research Part C: Emerging Technologies}, vol. 133, p. 103421, 2021.

\bibitem{wang2022distributed}
J.~Wang, Y.~Lian, Y.~Jiang, Q.~Xu, K.~Li, and C.~N. Jones, ``Distributed deep-lcc for cooperatively smoothing large-scale mixed traffic flow via connected and automated vehicles,'' \emph{arXiv preprint arXiv:2210.13171}, 2022.

\bibitem{shu2023safety}
Y.~Shu, J.~Zhou, and F.~Zhang, ``Safety-critical decision-making and control for autonomous vehicles with highest priority,'' in \emph{2023 IEEE Intelligent Vehicles Symposium (IV)}.\hskip 1em plus 0.5em minus 0.4em\relax IEEE, 2023, pp. 1--8.

\bibitem{shi2023deep}
H.~Shi, D.~Chen, N.~Zheng, X.~Wang, Y.~Zhou, and B.~Ran, ``A deep reinforcement learning based distributed control strategy for connected automated vehicles in mixed traffic platoon,'' \emph{Transportation Research Part C: Emerging Technologies}, vol. 148, p. 104019, 2023.

\bibitem{zhang2023privacy}
K.~Zhang, K.~Chen, Z.~Li, J.~Chen, and Y.~Zheng, ``Privacy-preserving data-enabled predictive leading cruise control in mixed traffic,'' \emph{IEEE Transactions on Intelligent Transportation Systems}, 2023.

\bibitem{ZHOU2024104885}
J.~Zhou and K.~Yang, ``A parameter privacy-preserving strategy for mixed-autonomy platoon control,'' \emph{Transportation Research Part C: Emerging Technologies}, vol. 169, p. 104885, 2024.

\bibitem{fang2024human}
Z.~Fang, J.~Wang, Z.~Wang, J.~Chen, G.~Yin, and H.~Zhang, ``Human--machine shared control for path following considering driver fatigue characteristics,'' \emph{IEEE Transactions on Intelligent Transportation Systems}, 2024.

\bibitem{zhou2024enhancing}
J.~Zhou, L.~Yan, and K.~Yang, ``Enhancing system-level safety in mixed-autonomy platoon via safe reinforcement learning,'' \emph{IEEE Transactions on Intelligent Vehicles}, pp. 1--13, 2024.

\bibitem{liang2024enhancing}
J.~Liang, K.~Yang, C.~Tan, J.~Wang, and G.~Yin, ``Enhancing high-speed cruising performance of autonomous vehicles through integrated deep reinforcement learning framework,'' \emph{arXiv preprint arXiv:2404.14713}, 2024.

\bibitem{shang2024decentralized}
X.~Shang, J.~Wang, and Y.~Zheng, ``Decentralized robust data-driven predictive control for smoothing mixed traffic flow,'' \emph{arXiv preprint arXiv:2401.15826}, 2024.

\bibitem{li2021reinforcement}
M.~Li, Z.~Cao, and Z.~Li, ``A reinforcement learning-based vehicle platoon control strategy for reducing energy consumption in traffic oscillations,'' \emph{IEEE Transactions on Neural Networks and Learning Systems}, vol.~32, no.~12, pp. 5309--5322, 2021.

\bibitem{liu2023safety}
H.~Liu, W.~Zhuang, G.~Yin, Z.~Li, and D.~Cao, ``Safety-critical and flexible cooperative on-ramp merging control of connected and automated vehicles in mixed traffic,'' \emph{IEEE Transactions on Intelligent Transportation Systems}, vol.~24, no.~3, pp. 2920--2934, 2023.

\bibitem{wang2022data}
J.~Wang, Y.~Zheng, Q.~Xu, and K.~Li, ``Data-driven predictive control for connected and autonomous vehicles in mixed traffic,'' in \emph{2022 American Control Conference (ACC)}.\hskip 1em plus 0.5em minus 0.4em\relax IEEE, 2022, pp. 4739--4745.

\bibitem{coulson2019data}
J.~Coulson, J.~Lygeros, and F.~D{\"o}rfler, ``Data-enabled predictive control: In the shallows of the deepc,'' in \emph{2019 18th European Control Conference (ECC)}.\hskip 1em plus 0.5em minus 0.4em\relax IEEE, 2019, pp. 307--312.

\bibitem{liu2024reinforcement}
L.~Liu, X.~Li, Y.~Li, J.~Li, and Z.~Liu, ``Reinforcement learning-based multi-lane cooperative control for on-ramp merging in mixed-autonomy traffic,'' \emph{IEEE Internet of Things Journal}, 2024.

\bibitem{chow2018risk}
Y.~Chow, M.~Ghavamzadeh, L.~Janson, and M.~Pavone, ``Risk-constrained reinforcement learning with percentile risk criteria,'' \emph{Journal of Machine Learning Research}, vol.~18, no. 167, pp. 1--51, 2018.

\bibitem{selim2022safe}
M.~Selim, A.~Alanwar, S.~Kousik, G.~Gao, M.~Pavone, and K.~H. Johansson, ``Safe reinforcement learning using black-box reachability analysis,'' \emph{IEEE Robotics and Automation Letters}, vol.~7, no.~4, pp. 10\,665--10\,672, 2022.

\bibitem{chow2018lyapunov}
Y.~Chow, O.~Nachum, E.~Duenez-Guzman, and M.~Ghavamzadeh, ``A lyapunov-based approach to safe reinforcement learning,'' \emph{Advances in neural information processing systems}, vol.~31, 2018.

\bibitem{cheng2019end}
R.~Cheng, G.~Orosz, R.~M. Murray, and J.~W. Burdick, ``End-to-end safe reinforcement learning through barrier functions for safety-critical continuous control tasks,'' in \emph{Proceedings of the AAAI conference on artificial intelligence}, vol.~33, no.~01, 2019, pp. 3387--3395.

\bibitem{emam2021safe}
Y.~Emam, G.~Notomista, P.~Glotfelter, Z.~Kira, and M.~Egerstedt, ``Safe reinforcement learning using robust control barrier functions,'' \emph{arXiv preprint arXiv:2110.05415}, 2021.

\bibitem{xiao2023barriernet}
W.~Xiao, T.-H. Wang, R.~Hasani, M.~Chahine, A.~Amini, X.~Li, and D.~Rus, ``Barriernet: Differentiable control barrier functions for learning of safe robot control,'' \emph{IEEE Transactions on Robotics}, 2023.

\bibitem{amos2017optnet}
B.~Amos and J.~Z. Kolter, ``Optnet: Differentiable optimization as a layer in neural networks,'' in \emph{International Conference on Machine Learning}.\hskip 1em plus 0.5em minus 0.4em\relax PMLR, 2017, pp. 136--145.

\bibitem{lindemann2023safe}
L.~Lindemann, M.~Cleaveland, G.~Shim, and G.~J. Pappas, ``Safe planning in dynamic environments using conformal prediction,'' \emph{IEEE Robotics and Automation Letters}, 2023.

\bibitem{fontana2023conformal}
M.~Fontana, G.~Zeni, and S.~Vantini, ``Conformal prediction: a unified review of theory and new challenges,'' \emph{Bernoulli}, vol.~29, no.~1, pp. 1--23, 2023.

\bibitem{yu2022surprising}
C.~Yu, A.~Velu, E.~Vinitsky, J.~Gao, Y.~Wang, A.~Bayen, and Y.~Wu, ``The surprising effectiveness of ppo in cooperative multi-agent games,'' \emph{Advances in Neural Information Processing Systems}, vol.~35, pp. 24\,611--24\,624, 2022.

\bibitem{ames2014control}
A.~D. Ames, J.~W. Grizzle, and P.~Tabuada, ``Control barrier function based quadratic programs with application to adaptive cruise control,'' in \emph{53rd IEEE Conference on Decision and Control}.\hskip 1em plus 0.5em minus 0.4em\relax IEEE, 2014, pp. 6271--6278.

\bibitem{aubin2011viability}
J.-P. Aubin, A.~M. Bayen, and P.~Saint-Pierre, \emph{Viability theory: new directions}.\hskip 1em plus 0.5em minus 0.4em\relax Springer Science \& Business Media, 2011.

\bibitem{feng2019string}
S.~Feng, Y.~Zhang, S.~E. Li, Z.~Cao, H.~X. Liu, and L.~Li, ``String stability for vehicular platoon control: Definitions and analysis methods,'' \emph{Annual Reviews in Control}, vol.~47, pp. 81--97, 2019.

\bibitem{liu2022structural}
D.~{}Liu, B.~Besselink, S.~Baldi, W.~Yu, and H.~L. Trentelman, ``On structural and safety properties of head-to-tail string stability in mixed platoons,'' \emph{IEEE Transactions on Intelligent Transportation Systems}, 2022.

\bibitem{wang2023deep}
J.~Wang, Y.~Zheng, K.~Li, and Q.~Xu, ``Deep-lcc: Data-enabled predictive leading cruise control in mixed traffic flow,'' \emph{IEEE Transactions on Control Systems Technology}, 2023.

\bibitem{wen2023modeling}
X.~Wen, S.~Jian, and D.~He, ``Modeling the effects of autonomous vehicles on human driver car-following behaviors using inverse reinforcement learning,'' \emph{IEEE Transactions on Intelligent Transportation Systems}, 2023.

\bibitem{vogel2003comparison}
K.~Vogel, ``A comparison of headway and time to collision as safety indicators,'' \emph{Accident analysis \& prevention}, vol.~35, no.~3, pp. 427--433, 2003.

\bibitem{das2019defining}
S.~Das and A.~K. Maurya, ``Defining time-to-collision thresholds by the type of lead vehicle in non-lane-based traffic environments,'' \emph{IEEE Transactions on Intelligent Transportation Systems}, vol.~21, no.~12, pp. 4972--4982, 2019.

\bibitem{xiao2021high}
W.~Xiao and C.~Belta, ``High-order control barrier functions,'' \emph{IEEE Transactions on Automatic Control}, vol.~67, no.~7, pp. 3655--3662, 2021.

\bibitem{zhou2022safety}
J.~Zhou and H.~Yu, ``Safety critical control of mixed-autonomy traffic via a single autonomous vehicle,'' in \emph{2022 IEEE 25th International Conference on Intelligent Transportation Systems (ITSC)}.\hskip 1em plus 0.5em minus 0.4em\relax IEEE, 2022, pp. 3089--3094.

\bibitem{jiang2001full}
R.~Jiang, Q.~Wu, and Z.~Zhu, ``Full velocity difference model for a car-following theory,'' \emph{Physical Review E}, vol.~64, no.~1, p. 017101, 2001.

\bibitem{fairclough1997effect}
S.~H. Fairclough, A.~J. May, and C.~Carter, ``The effect of time headway feedback on following behaviour,'' \emph{Accident Analysis \& Prevention}, vol.~29, no.~3, pp. 387--397, 1997.

\bibitem{xiao2021adaptive}
W.~Xiao, C.~Belta, and C.~G. Cassandras, ``Adaptive control barrier functions,'' \emph{IEEE Transactions on Automatic Control}, vol.~67, no.~5, pp. 2267--2281, 2021.

\end{thebibliography}
\vspace{11pt}
\begin{IEEEbiography}
[{\includegraphics[width=1in,height=1.25in,clip,keepaspectratio]{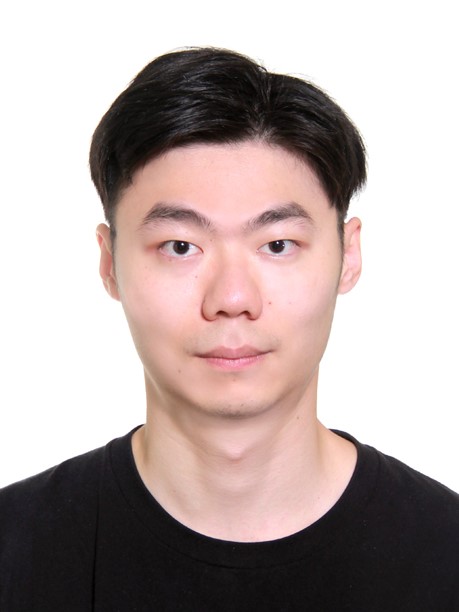}}]{Jingyuan Zhou} received the B.Eng. degree in Electronic Information Science and Technology from Sun Yat-sen University, Guangzhou, China, in 2022. He is currently working towards a Ph.D. degree with the National University of Singapore. His research interests include safe and secure control of connected and automated vehicles in intelligent transportation systems.
\end{IEEEbiography}
\vspace{11pt}
\begin{IEEEbiography}
[{\includegraphics[width=1in,height=1.25in,clip,keepaspectratio]{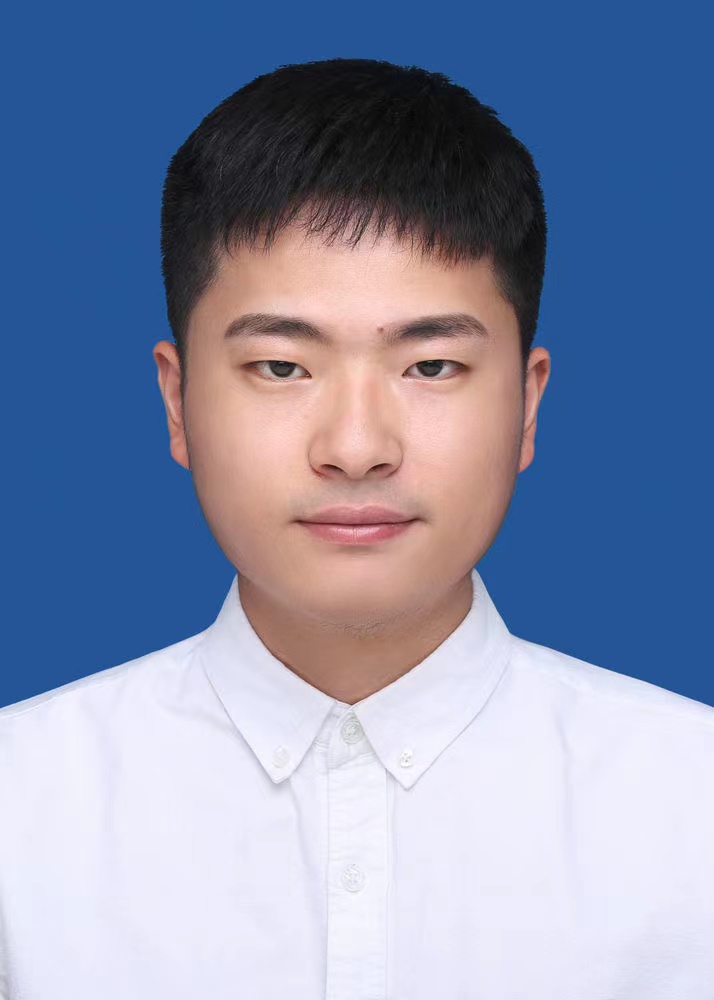}}]{Longhao Yan} received the B.Eng. degree and M.Eng. degree in School of Electronics and Control Engineering from Chang’an University, Xi’an, China, in 2019 and 2022 respectively. He is currently working towards a Ph.D. degree with the National University of Singapore. His research interests include lateral control and trajectory prediction of intelligent transportation systems.
\end{IEEEbiography}
\vspace{11pt}
\begin{IEEEbiography}
[{\includegraphics[width=1in,height=1.25in,clip,keepaspectratio]{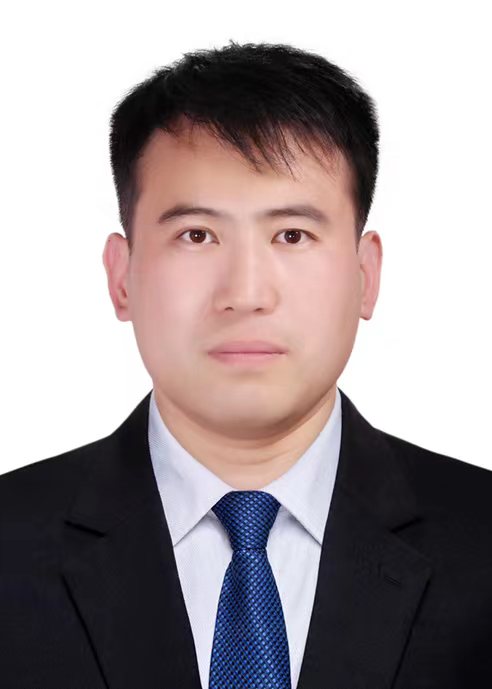}}]{Jinhao Liang}
received the B.S. degree from School of Mechanical Engineering, Nanjing University of Science and Technology, Nanjing, China, in 2017, and Ph.D. degree from School of Mechanical Engineering, Southeast University, Nanjing, China, in 2022. Now he is a Research Fellow with Department of Civil and Environmental Engineering, National University of Singapore. His research interests include vehicle dynamics and control, autonomous vehicles, and vehicle safety assistance system.
\end{IEEEbiography}
\vspace{11pt}
\begin{IEEEbiography}
[{\includegraphics[width=1in,height=1.25in,clip,keepaspectratio]{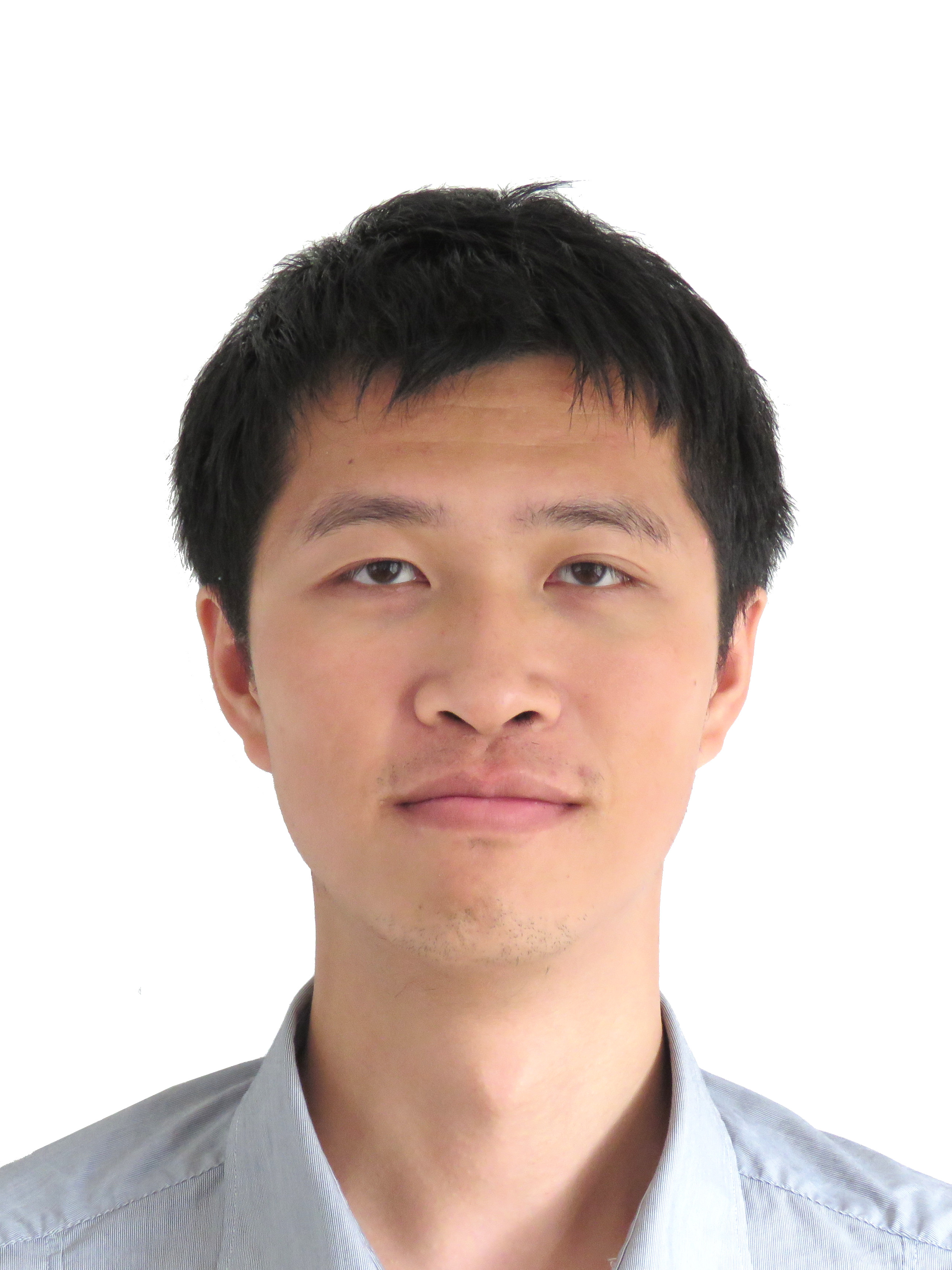}}]{Kaidi Yang}  is an Assistant Professor in the Department of Civil and Environmental Engineering at the National University of Singapore. Prior to this, he was a postdoctoral researcher with the Autonomous Systems Lab at Stanford University. He obtained a PhD degree from ETH Zurich and M.Sc. and B.Eng. degrees from Tsinghua University. His main research interest is the operation of future mobility systems enabled by connected and automated vehicles (CAVs) and shared mobility.
\end{IEEEbiography}
\vfill

\end{document}